\documentclass[manuscript]{aastex}

\usepackage[]{amsmath}

\received{receipt date}
\revised{revision date}
\accepted{acceptance date}
\shorttitle{kappa distribution}
\shortauthors{Bautista \& Mendoza}

\begin{document}

\title{Testing the existence of non-Maxwellian electron distributions in
\ion{H}{2} regions after assessing atomic data accuracy}

\author{C. Mendoza\altaffilmark{1}}
\affil{Department of Physics, Western Michigan University,
Kalamazoo, MI 49008, USA}
\altaffiltext{1}{Permanent address: Centro de F\'isica, Instituto Venezolano de Investigaciones
Cient\'ificas (IVIC), PO Box 20632, Caracas 1020A,  Venezuela}
\email{claudio.mendozaguardia@wmich.edu}
\and
\author{M.~A. Bautista}
\affil{Department of Physics, Western Michigan University,
Kalamazoo, MI 49008, USA}
\email{manuel.bautista@wmich.edu}


\begin{abstract}
  The classic optical nebular diagnostics [\ion{N}{2}], [\ion{O}{2}], [\ion{O}{3}], [\ion{S}{2}], [\ion{S}{3}], and [\ion{Ar}{3}] are employed to search for evidence of non-Maxwellian electron distributions, namely $\kappa$~distributions, in a sample of well-observed Galactic \ion{H}{2} regions. By computing new effective collision strengths for all these systems and $A$-values when necessary (e.g. \ion{S}{2}), and by comparing with previous collisional and radiative datasets, we have been able to obtain realistic estimates of the electron-temperature dispersion caused by the atomic data, which in most cases are not larger than $\sim 10$\%. If the uncertainties due to both observation and atomic data are then taken into account, it is plausible to determine for some nebulae a representative average temperature while in others there are at least two plasma excitation regions. For the latter, it is found that the diagnostic temperature differences in the high-excitation region, e.g. $T_e$(\ion{O}{3}), $T_e$(\ion{S}{3}), and $T_e$(\ion{Ar}{3}), cannot be conciliated by invoking $\kappa$~distributions. For the low excitation region, it is possible in some, but not all, cases to arrive at a common, lower temperature for [\ion{N}{2}], [\ion{O}{2}], and [\ion{S}{2}] with $\kappa\approx 10$, which would then lead to significant abundance enhancements for these ions. An analytic formula is proposed to generate accurate  $\kappa$-averaged excitation rate coefficients (better than 10\% for $\kappa \geq 5$) from temperature tabulations of the Maxwell--Boltzmann effective collision strengths.
\end{abstract}



\normalsize


\section{Introduction}

A fundamental and still unresolved problem in the understanding of astronomical photoionized plasmas, particularly \ion{H}{2} regions and planetary nebulae, is the systematic higher metal abundances derived from recombination lines (RL) relative to those from collisionally excited lines (CEL). The magnitude of these discrepancies varies from a few percent up to puzzling factors as large as 70 \citep[see, for instance,][]{liu00, liu01}. Several plausible explanations have been proposed to account for their cause, among them temperature fluctuations \citep{pei67}, hydrogen deficient inclusions \citep{liu00, sta07}, and dense clumps ionized by X rays \citep{erc09}. However, all these hypotheses present certain limitations and no conclusive evidence has been put forward to either prove or dispute them.

\citet{nic12} have suggested that the apparent temperature variations in nebular plasmas are the result of electron velocity distributions that deviate from the generally assumed Maxwell--Boltzmann (MB), proposing the $\kappa$~distribution as a more functional alternative. It is therein argued that excitation rates sufficiently boosted from the MB values are obtained with $10\leq \kappa\leq 20$, implying that the traditional temperature diagnostics from forbidden line ratios systematically overestimate the true kinetic temperature; therefore, ion abundances derived from CEL would be underestimated. In their analysis, two crucial approximations are made: (i) electron impact collision strengths are constant with energy, which in  practice is not the case due to strong resonance contributions; and (ii) the effects of the $\kappa$~distribution on the de-excitation rate coefficients are ignored.

\citet{bin12} have found that the \ion{O}{3} temperatures from CEL are consistently higher than those of \ion{S}{3} in an extensive group of both Galactic and extragalatic \ion{H}{2} regions with gas metallicities higher than 0.2 solar. Among the causes for this excess, the $\kappa$~distribution has again been invoked. \citet{nic13} have continued their work on the $\kappa$~distributions by revising the effects of different datasets of effective collision strengths on the electron temperatures obtained from nebular diagnostics such as [\ion{N}{2}], [\ion{O}{2}], [\ion{O}{3}], [\ion{S}{2}], and [\ion{S}{3}]. In particular, by comparing the \ion{O}{3} electron temperature obtained using recent collision strengths \citep{pal12} with those derived from earlier datasets, they have concluded that the previous electron temperatures have been seriously overestimated. Since any spectroscopic evidence of non-Maxwellian electron distributions must rely on accurate atomic data, it is indispensable to verify this claim. It must be added that \citet{nic13} did not consider the accuracy of the radiative rates ($A$-values) which is also a source of uncertainty in spectral diagnostics. For a full treatment of the propagation of atomic data uncertainties in spectral diagnostics, see \citet{bau13}.

More recently, \citet{dop13} have made an effort to estimate both MB and $\kappa$-averaged collisional rates from energy tabulations of raw collision strengths, and stress the need for a systematic inventory and assessment of the newer atomic data and their effects on photoionization models before really establishing the prevalence of the $\kappa$~distribution in nebular plasmas. This is in fact the main motive of the present work. We make use of the familiar nebular diagnostics---namely [\ion{N}{2}], [\ion{O}{2}], [\ion{O}{3}],  [\ion{S}{2}], [\ion{S}{3}], and [\ion{Ar}{3}]---to search for evidence of $\kappa$~distributions in well-observed \ion{H}{2} regions. In Section~\ref{rates} we describe the effects of the $\kappa$~distribution on both the electron excitation and de-excitation rate coefficients for forbidden transitions. Since energy-tabulated electron collision strengths with at least a similar degree of reliability are required in this analysis, we decided to recalculate them from scratch (Section~\ref{upsilon}). The sensitivity of the electron temperature to the $A$-values is also evaluated in Section~\ref{aval}. With the new data, we derive in Section~\ref{h2temp} electron temperatures for the \ion{H}{2} regions studied by \citet{nic12} and compare them with those obtained with representative effective collision strengths published in the past two decades. The possibilities of the $\kappa$~distribution in reducing CEL temperatures and in conciliating the temperature differences in plasma regions are also investigated (Section~\ref{kappaeffects}). Conclusions are finally discussed in Section~\ref{summary}.


\section{$\kappa$-averaged rates}
\label{rates}

Spectral emission of non-Maxwellian plasmas has been studied in detail by \citet{bry05} from which we recapitulate the formalism for electron impact excitation and de-excitation within the $\kappa$-distribution framework. For an ionic transition between levels $i\rightarrow j$ induced by electron impact, the excitation and de-excitation generalized rate coefficients may be written as
\begin{equation}\label{excitate}
C_{ij}=\frac{2\sqrt{\pi}\alpha ca_0^2(R/k_B)^{1/2}}{g_i T^{1/2}}
\exp{\left(-\frac{\Delta E_{ij}}{k_BT}\right)}\Upsilon_{ij}
\end{equation}
and
\begin{equation}\label{deexcitate}
C_{ji}=\frac{2\sqrt{\pi}\alpha ca_0^2(R/k_B)^{1/2}}{g_j T^{1/2}}\Upsilon_{ji}
\end{equation}
where $\Delta E_{ij}$ is the energy separation between the two levels and $g_i$ and $g_j$ are their respective statistical weights. The other quantities are the standard fine structure constant, $\alpha$, speed of light, $c$, Bohr radius, $a_0$, Rydberg constant, $R$, and Boltzmann constant, $k_B$. For a particular normalized electron-energy distribution function $f(E)$, the kinetic temperature $T\equiv 2\overline{E}/3k_B$ is defined in terms of the mean energy
\begin{equation}
\overline{E} =\int Ef(E)\mathrm{d}E\ .
\end{equation}
The effective collision strengths for excitation, $\Upsilon_{ij}$, and de-excitation, $\Upsilon_{ji}$, involve energy averages over the quantum mechanical cross section $\sigma(E)$ for the transition, which may be conveniently expressed in terms of the symmetric collision strength
\begin{equation}
           \Omega_{ij}(E)=g_i \left(\frac{E_i}{R}\right)\left(\frac{\sigma_{ij}(E_i)}{\pi a_0^2}\right)
           =g_j \left(\frac{E_j}{R}\right)\left(\frac{\sigma_{ji}(E_j)}{\pi a_0^2}\right)
           \end{equation}
where $E_i$ and $E_j$ are the free-electron energies relative to the $i$th and $j$th levels, respectively. They are then expressed through the integrals
\begin{equation}
\Upsilon_{ij}=\frac{\sqrt{\pi}}{2}\exp{\left(\frac{\Delta E_{ij}}{k_BT}\right)}
\int_0^\infty\Omega_{ij}(E_i)\left(\frac{E_i}{k_BT}\right)^{-1/2}f(E_i)
\mathrm{d}E_i
\end{equation}
and
\begin{equation}
\Upsilon_{ji}=\frac{\sqrt{\pi}}{2}\int_0^\infty\Omega_{ij}(E_j)\left(\frac{E_j}
{k_BT}\right)^{-1/2}f(E_j)\mathrm{d}E_j\ .
\end{equation}

For the MB electron distribution
\begin{equation}
f_{T_e}(E) =\frac{2}{\sqrt{\pi}k_BT_e}\left(\frac{E}{k_BT_e}\right)^{1/2}
\exp{\left(-\frac{E}{k_BT_e}\right)}\ ,
\end{equation}
$T$ becomes the familiar electron temperature $T_e$ and the effective collision strengths for excitation and de-excitation are also symmetric
\begin{equation}\label{ups}
\Upsilon_{ij}=\Upsilon_{ji}=\Upsilon_{ji}^{\rm MB}(T_e)=\int_0^\infty \Omega_{ij}(E_j)\exp{\left(-\frac{E_j}{k_BT_e}\right)}\,
 \mathrm{d}\left(\frac{E_j}{k_BT_e}\right)\ ,
\end{equation}
an unmistakeable signature of detailed balance for this distribution.

On the other hand, the $\kappa$~distribution of electron energies
\begin{equation}\label{kappa}
f_{\kappa,E_\kappa}(E) =\frac{2\kappa^{-3/2}}{\sqrt{\pi}E_\kappa}\left(\frac{E}{E_\kappa}\right)^{1/2}
\frac{\Gamma(\kappa+1)}{\Gamma(\kappa-1/2)}
\left(1+\frac{E}{\kappa E_\kappa}\right)^{-(\kappa+1)}
\end{equation}
has a characteristic energy $E_\kappa$ that is related to the kinetic temperature
\begin{equation}
E_\kappa = k_BT_\kappa(\kappa-3/2)/\kappa
\end{equation}
 where the $\kappa$ parameter ($3/2\leq\kappa\leq\infty$) gives a measure of the deviation from the MB distribution, converging to the latter as $\kappa\rightarrow\infty$. Detailed balance is now broken and the asymmetric effective collision strengths are given by
\begin{multline}
\label{kappa_excitation}
\Upsilon_{ij}^\kappa(T_\kappa) = (\kappa-3/2)^{-3/2}
\frac{\Gamma(\kappa+1)}{\Gamma(\kappa -1/2)}\exp{\left(\frac{\Delta E_{ij}}{k_BT_\kappa}\right)}\\
\times\int_0^\infty\Omega_{ij}\left(1+\frac{E_j+\Delta E_{ij}}
{(\kappa-3/2)k_BT_\kappa}\right)^{-(\kappa+1)}\mathrm{d}\left(\frac{E_j}{k_BT_\kappa}\right)
\end{multline}
and
\begin{equation}
\Upsilon_{ji}^\kappa(T_\kappa) = (\kappa-3/2)^{-3/2}\frac{\Gamma(\kappa+1)}{\Gamma(\kappa -1/2)}
\int_0^\infty\Omega_{ij}\left(1+\frac{E_j}
{(\kappa-3/2)k_BT_\kappa}\right)^{-(\kappa+1)}\mathrm{d}\left(\frac{E_j}{k_BT_\kappa}\right)\ .
\end{equation}

\begin{figure}
  \rotatebox{0}{\resizebox{\hsize}{\hsize}{\plotone{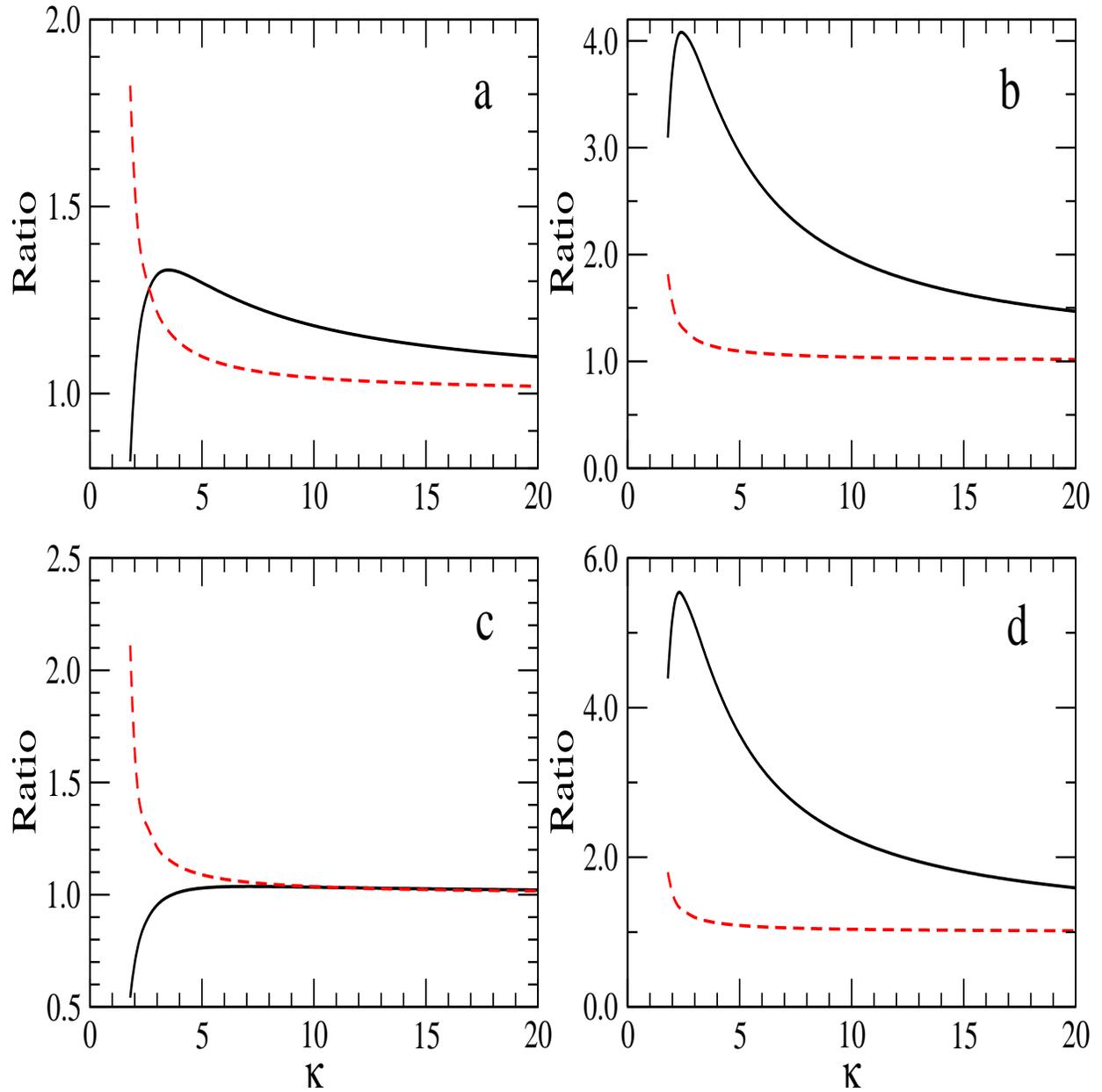}}}
  \caption{Ratio of the effective collision strengths $\Upsilon^\kappa/\Upsilon^{\rm MB}$ at $T=10^4$~K as a function of $\kappa$ for electron impact excitation (solid line) and de-excitation (dashed line) of the transitions:
  (a) [\ion{O}{2}] ${\rm 2s^22p^3\ ^4S^o_{3/2}}-{\rm ^2D^o_{3/2}}$; (b) [\ion{O}{2}] ${\rm 2s^22p^3\ ^4S^o_{3/2}}-{\rm ^2P^o_{1/2}}$; (c) [\ion{O}{3}] ${\rm 2s^22p^2\ ^3P_0}-{\rm ^1D_2}$; and (d) [\ion{O}{3}] ${\rm 2s^22p^2\ ^3P_0}-{\rm ^1S_0}$. \label{ratio_kappa}
  }

\end{figure}

\begin{figure}
   \rotatebox{0}{\resizebox{\hsize}{\hsize}{\plotone{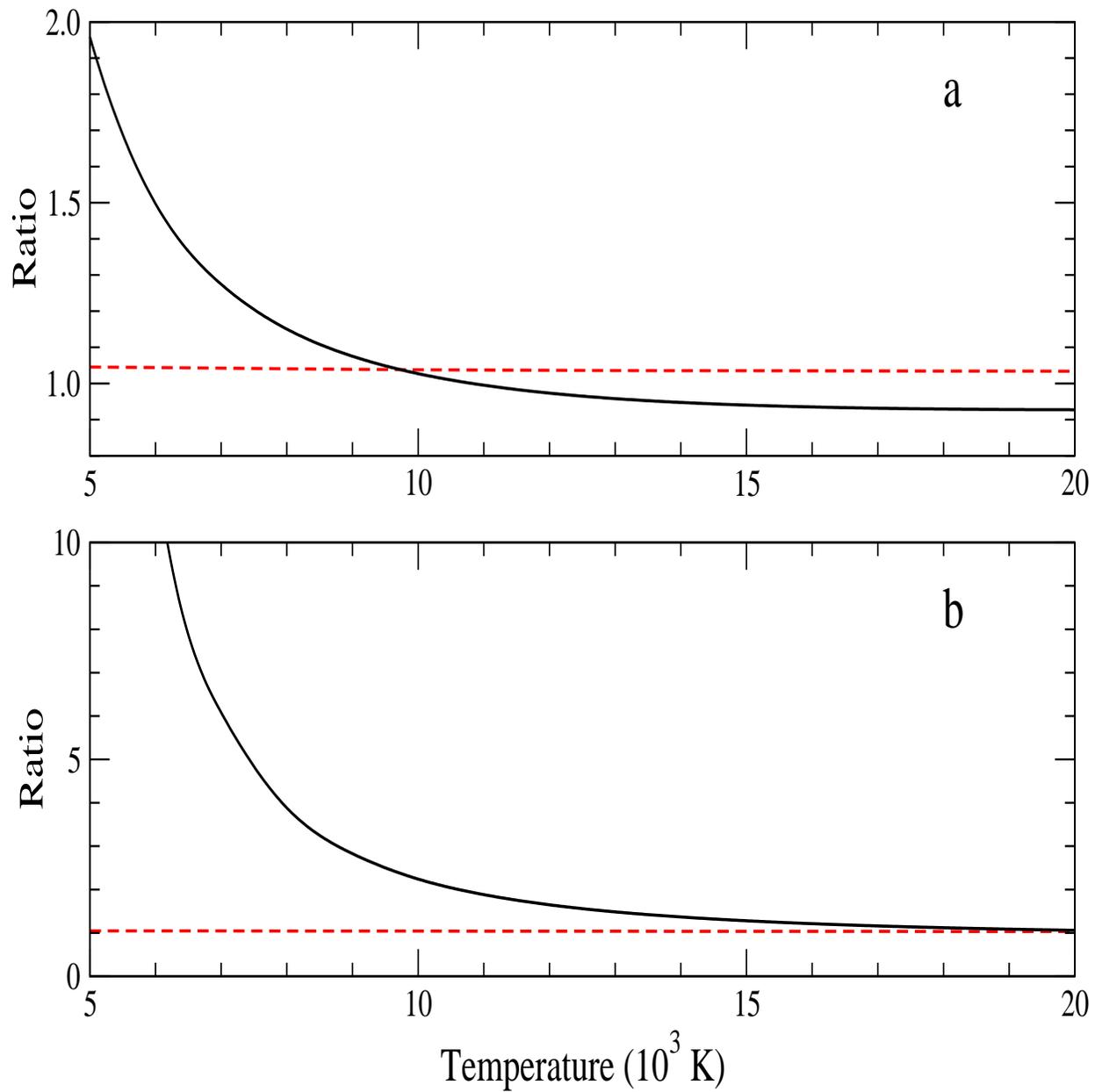}}}
  \caption{Ratio of the effective collision strengths $\Upsilon^\kappa/\Upsilon^{\rm MB}$ at $\kappa=10$ as a function of temperature for electron impact excitation (solid line) and de-excitation (dashed line) of the [\ion{O}{3}] transitions: (a) ${\rm 2s^22p^2\ ^3P_0}-{\rm ^1D_2}$ and (b) ${\rm 2s^22p^2\ ^3P_0}-{\rm ^1S_0}$. \label{ratio_temp}
  }
\end{figure}

\begin{figure}
   \rotatebox{0}{\resizebox{\hsize}{\hsize}{\plotone{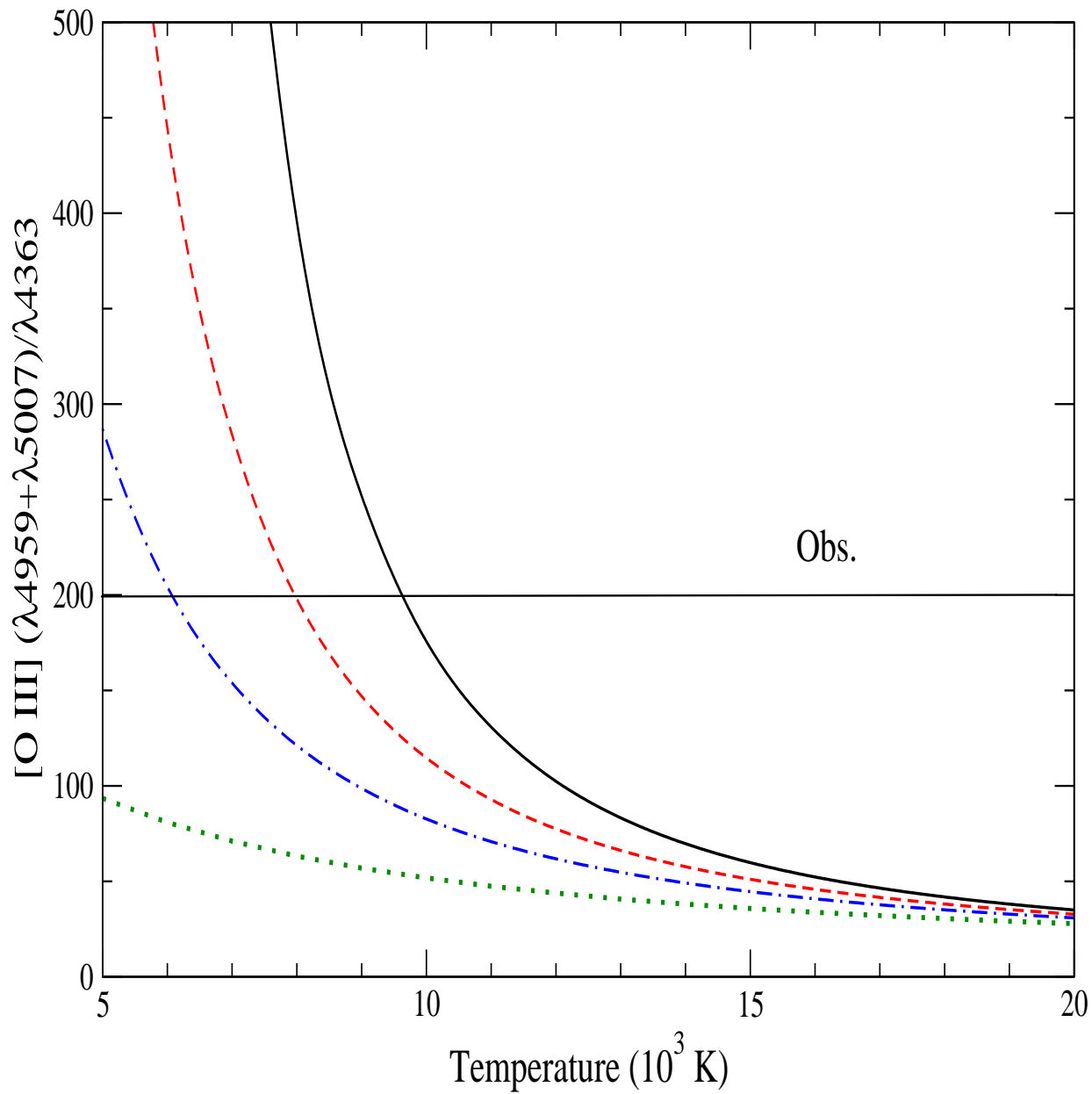}}}
  \caption{[\ion{O}{3}] $({\lambda}4959+{\lambda}5007)/{\lambda}4363$ line ratio as a function of temperature at
  $N_e=1.5\times10^4$~cm$^{-3}$ for MB and $\kappa$~distributions. Solid curve, MB distribution. Dashed curve, $\kappa=20$. Dash-dotted curve, $\kappa=10$. Dotted curve, $\kappa=5$. An observed ratio of 200 is indicated. \label{o3_line_ratio}
  }
\end{figure}

In Figure~\ref{ratio_kappa}, the ratio $\Upsilon^\kappa/\Upsilon^{\rm MB}$ is plotted as a function of $\kappa$ for the forbidden transitions arising from the ground states of \ion{O}{2} and \ion{O}{3}. It may be seen that the $\kappa$~distribution increases the de-excitation effective collision strength by up to a factor of two for $\kappa\lesssim 5$, while for higher $\kappa$ the enhancements are marginal. For excitation, the $\kappa$~distribution causes large differences for $\kappa\lesssim 5$, in particular increments as large as a factor of six for the transitions with the larger $\Delta E_{ij}$; for such transitions, they remain sizeable ($>50$\%) even for $\kappa> 20$.
The behavior of this ratio with temperature in \ion{O}{3} is depicted in Figure~\ref{ratio_temp}, where it may be appreciated that, for de-excitation, the ratio is temperature independent and not larger than 10\%; for excitation, on the other hand, exponential increases are found with decreasing temperature for $T< 10^4$~K especially for the transition ${^3\rm P}_0-{^1\rm S}_0$ due to its comparatively larger $\Delta E_{ij}$. In Figure~\ref{o3_line_ratio}, the [\ion{O}{3}] $({\lambda}4959 +{\lambda}5007)/{\lambda}4363$ line ratio is plotted as a function of temperature for both the MB and $\kappa$ distributions at an electron density of $N_e=1.5\times10^4$~cm$^{-3}$. It is therein shown that, for an observed line ratio of 200 say, MB results in $T_e=9700$~K while the $\kappa$~distribution gives significantly lower values: $T_\kappa=8000$~K and $T_\kappa=6100$~K for $\kappa=20$ and $\kappa=10$, respectively.

These findings are in general agreement with \citet{nic12}, and support their approximation of neglecting the contributions of the $\kappa$~distribution to the de-excitation rate coefficient. Regarding their assumption of estimating the $\kappa$-averaged excitation rate coefficient with a constant collision strength, we have found that a better choice would be to derive it from the MB temperature-dependent effective collision strength for the transition through a simplified version of equation~(\ref{kappa_excitation}), namely
\begin{multline}
\Upsilon_{ij}^\kappa(T) \approx (\kappa-3/2)^{-3/2}
\frac{\Gamma(\kappa+1)}{\Gamma(\kappa -1/2)}\exp{\left(\frac{\Delta E_{ij}}{k_BT}\right)}\,\Upsilon_{ji}^{\rm MB}(T)\\
\times\int_0^\infty\left(1+\frac{E_j+\Delta E_{ij}}
{(\kappa-3/2)k_BT}\right)^{-(\kappa+1)}\mathrm{d}\left(\frac{E_j}{k_BT}\right)\ ,
\end{multline}
which can be integrated analytically to give
\begin{equation}
\label{approx_kappa}
\Upsilon_{ij}^\kappa(T) \approx \frac{(\kappa-3/2)^{-1/2}}{\kappa}
\frac{\Gamma(\kappa+1)}{\Gamma(\kappa -1/2)}
\left(1+\frac{\Delta E_{ij}}{(\kappa-3/2)k_BT}\right)^{-\kappa}
\exp{\left(\frac{\Delta E_{ij}}{k_BT}\right)}\,\Upsilon_{ji}^{\rm MB}(T)\ .
\end{equation}
Equation~(\ref{approx_kappa}) is a small yet significant improvement over equation~(16) in \citet{nic12} inasmuch as now the MB effective collision strength $\Upsilon_{ji}^{\rm MB}(T)$ is temperature dependent rather than constant, and enables the accurate determination of $\kappa$-averaged excitation rate coefficients (better than 10\% for $\kappa \geq 5$) from temperature tabulations of the MB effective collision strengths. Furthermore, as mentioned in \citet{nic12}, the key term in equation~(\ref{approx_kappa}) is $T/T_{\rm ex}$, the ratio of the kinetic temperature relative to the excitation temperature $T_{\rm ex}\equiv \Delta E_{ij}/k_B$, since $\kappa$-distribution enhancements to the excitation rate coefficient become conspicuous for $T/T_{\rm ex}\ll 1$. Hence, IR transitions within the ionic ground term, where usually  $T_{\rm ex}< 10^3$~K, are not expected to show noticeable departures from MB conditions in nebular plasmas.

The optical diagnostics we study in the present work are those widely used in observational work to determine nebular plasma properties---namely [\ion{N}{2}], [\ion{O}{2}], [\ion{O}{3}], [\ion{S}{2}], [\ion{S}{3}], and [\ion{Ar}{3}]---the specific line ratios being listed in Table~\ref{lineratios}. Since the collision strengths for these transitions are not readily available from the original sources, and the present analysis requires at least an even degree of reliability, we had no alterative but to recalculate them. [They are openly available from the AtomPy\footnote{\tt http://bit.ly/K5oAfD} atomic data curation service on Google Drive; for an ion identified by the tuple $(zz,nn)$, where $zz$ (two-character string) and $nn$ (two-character string) stand for the atomic and electron numbers, the spreadsheet containing level energies, $A$-values, collision strengths, and effective collision strengths is located at {\tt IsonuclearSequences/zz/zz\_nn}.]


\section{Collision strengths}
\label{upsilon}

 When computing effective collision strengths for both the MB and $\kappa$ distributions used in Figures~\ref{ratio_kappa}--\ref{ratio_temp}, it is important to take into account that the collision strengths for forbidden transitions are dominated by dense packs of resonances in the region near the excitation threshold; therefore, it is the mesh between $E_j=0$ (threshold) and $\sim 1$~Ryd that determines their value at typical nebular temperatures ($T_e\sim 10^4$~K).

 In the present work, collision strengths for the forbidden transitions within the ground configuration of the ions specified in Table~\ref{lineratios} are computed with the Breit--Pauli $R$-matrix method \citep{ber78, ber87, sco80, sco82}. Multi-configuration wave functions for the target representations have been obtained in a Thomas--Fermi--Dirac potential with the atomic structure code {\sc autostructure}\footnote{\tt http://amdpp.phys.strath.ac.uk/autos/} \citep{eis74, bad86, bad97}. The scattering calculations are carried out in $LS$-coupling taking into account partial wave contributions with $L\leq 9$ and including the non-fine-structure mass, velocity, and Darwin one-body relativistic corrections. The intermediate coupling (IC) collision strengths for fine-structure levels are obtained by means of the Intermediate Coupling Frame Transformation (ICFT) method of \citet{gri98} using energy meshes with a resolution of $10^{-4}$~Ryd for most ions except \ion{Ar}{3} where a finer step of $5\times 10^{-5}$~Ryd was required. The ICFT method is computationally less demanding than a fully relativistic Breit--Pauli calculation without compromising accuracy, thus allowing the handling of more complex target representations. The ionic targets are specified in Table~\ref{targets} and will be henceforth briefly described (Sections~\ref{n2}--\ref{ar3}). The resulting effective collision strengths are also compared with other representative datasets in order to estimate their accuracy. In this respect, since one-to-one comparisons are usually hindered by the diverse temperature tabulations found in publication, a useful measure (see Table~\ref{ups}) is the scatter of the effective collision strengths at $T_e=10^4$~K for the leading transitions that populate the diagnostic upper levels (this temperature, representative of the order of nebular temperatures, is always given). As shown in Table~\ref{ups}, the mean dispersions are not greater than 15\%. 


\subsection{\ion{N}{2}}
\label{n2}

\begin{figure}
   \rotatebox{0}{\resizebox{\hsize}{\hsize}{\plotone{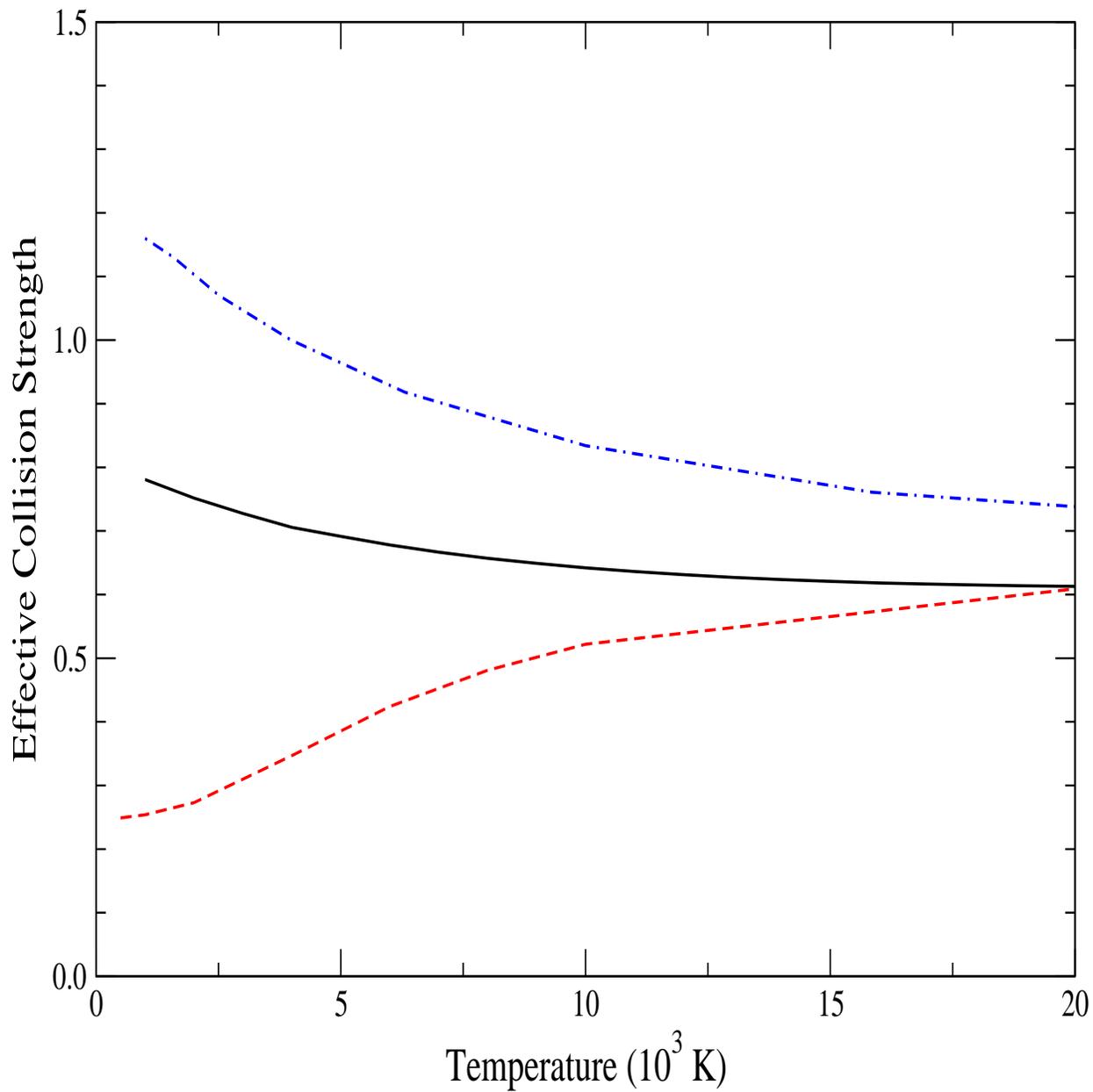}}}
  \caption{Effective collision strengths for the transition [\ion{N}{2}] ${\rm ^1D_2}-{\rm ^1S_0}$. Solid curve, present $R$-matrix calculation. Dashed curve, 58-level IC $R$-matrix calculation \citep{tay11}. Dash-dotted curve, 12-term $LS$ $R$-matrix calculation \citep{len94} \label{N2_ups}.
  }
\end{figure}

As indicated in Table~\ref{targets}, our \ion{N}{2} target representation includes in the close-coupling expansion the ten lowest $LS$ terms and configuration interaction within complexes with principal quantum number $n\leq 3$. The resulting effective collision strengths are compared with two other datasets: (i) the $R$-matrix calculation by \citet{len94} with a 12-term target representation in an $LS$-coupling framework, the IC collision strengths being obtained by algebraic recoupling of the $LS$ reactance matrices; and (ii) the IC $B$-spline, Breit--Pauli $R$-matrix method of \citet{tay11} with a target containing 58 fine-structure levels. The present scheme is similar to that by \citet{len94}, but our IC collision strengths are obtained with the ICFT method based on multi-channel quantum defect theory \citep{gri98}. The agreement with \citet{tay11} is in general within 10\% except for the ${\rm 2s^22p^2\ ^1D_2}-{\rm ^1S_0}$ transition at the lower temperatures ($T_e< 10^4$~K) where larger discrepancies are found (see Fig.~\ref{N2_ups}). Significant differences ($\sim 20\%$) are also found with the data of \citet{len94} for this same transition as shown in Fig.~\ref{N2_ups} and for ${\rm ^3P_2}-{\rm ^3P_1}$.

The outcome of this comparison, i.e. good agreement for most transitions but noticeable discrepancies for a selected few, is typical of collisional rates dominated by resonances, where small variations in resonance patterns can cause unexpected effects. Nonetheless, the discrepancies for the ${\rm ^3P}_J-{\rm ^3P}_{J'}$ transitions have little impact on the optical emission spectra of typical nebulae since the fine-structure levels within the ground multiplet are nearly thermalized at electron densities of $N_e\sim 10^4$~cm$^{-3}$ and temperatures of $T_e\sim 10^4$~K, while the optical transitions arising from the first and second excited terms are dominated by excitations from the ground multiplet.


\subsection{\ion{O}{2}}

Effective collision strengths for the forbidden transitions in \ion{O}{2} have been recently assessed in Appendix~A of \citet{sta12}, where the most elaborate Breit--Pauli $R$-matrix calculation by \citet{kis09} (21-level target) is in very good accord (a few percent) with \citet{mon06, pra06} (16-level target, Breit--Pauli $R$-matrix) and \citet{tay07} (62-level target, $B$-spline, Breit--Pauli $R$-matrix) except for the ${2s^22p^3\ ^2D^o_{3/2}}-{^2D^o_{5/2}}$ transition; for this transition, differences at the lower temperatures are around 36\% and 25\%, respectively. Notoriously larger discrepancies are found with the data by \citet{mcl98}; hence, they will not be further considered in the present analyses.

Our target expansion includes 19 $LS$ terms in the close-coupling expansion and correlation configurations with $n\leq 3$ (see Table~\ref{targets}). Our resulting effective collision strengths are found to be in general around 25\% higher than those of \citet{kis09} which are believed to be due to our smaller target representation.


\subsection{\ion{O}{3}}

The level of agreement between the effective collision strengths computed for \ion{O}{3} by \citet{len94}, \citet{agg99}, and \citet{pal12} is unimpressively $\sim 20\%$. \citet{len94} computed their $R$-matrix data in $LS$-coupling in a 12-term target approximation, obtaining IC effective collision strengths by algebraic recoupling of the reactance matrix. The data by \citet{agg99} were computed in a similar fashion but with a 26-term target representation. The recent calculation by \citet{pal12} was performed with an IC Breit--Pauli $R$-matrix method with a 19-level target and extensive configuration interaction within $n\leq 4$ complexes. This approach implements experimental thresholds, and also includes in the relativistic Hamiltonian the two-body Breit terms that are usually neglected in relativistic scattering work.

As shown in Table~\ref{targets}, the present $R$-matrix calculation has been carried out with a model target of 9 $LS$ terms and extensive configuration interaction within the $n\leq 3$ complexes. The resulting effective collision strengths are found to be around 30\% below those of \citet{pal12} except for the ${\rm 2s^22p^2\ ^1D_2}-{\rm ^1S_0}$ quadrupole transition where this trend is reversed. Such differences are perhaps caused by our shorter close-coupling expansion.


\subsection{\ion{S}{2}}

\begin{figure}
   \rotatebox{0}{\resizebox{\hsize}{\hsize}{\plotone{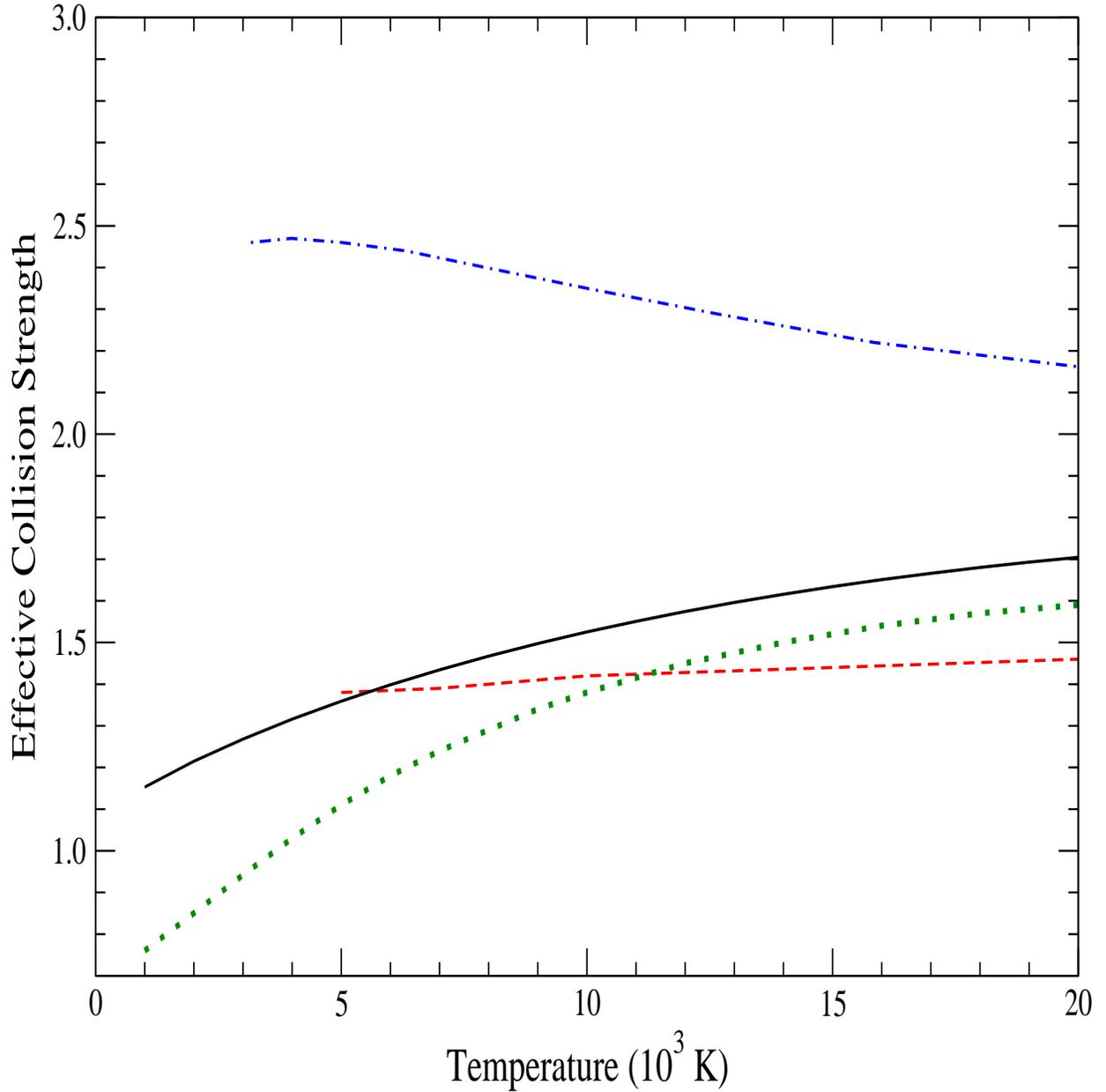}}}
  \caption{Effective collision strengths for the transition [\ion{S}{2}] ${\rm ^4S^o_{3/2}}-{\rm ^2P^o_{3/2}}$. Solid curve, present $R$-matrix calculation. Dashed curve, 70-level IC $R$-matrix calculation \citep{tay10}. Dash-dotted curve, 18-term $LS$ $R$-matrix calculation \citep{ram96}. Dotted curve, 12-term $LS$ $R$-matrix calculation \citep{cai93} \label{S2_ups}.
  }
\end{figure}

\ion{S}{2} has been the most difficult of all the systems treated here. We have explored several configuration expansions including orbitals up to $n=4$ and $n=5$. It is found that the best representation is obtained with a relatively small expansion that allows for single electron promotions from the 2p sub-shell and including in the close-coupling expansion the lowest 17 $LS$ terms (see Table~\ref{targets}). Previous $R$-matrix work has been carried out by \citet{cai93} and \citet{ram96} with target expansions including the lower 12 and 18 $LS$ terms, respectively; the IC effective collision strengths were obtained by algebraic recoupling of the reactance matrices. \citet{tay10} have performed an IC $B$-spline, Breit--Pauli $R$-matrix calculation with a 70-level target. In general, we find good agreement (within $\sim 20\%$) with \citet{tay10} but discrepancies around a factor of two for the ${\rm 3s^23p^3\ ^4S^o_{3/2}-{^2P}^o_J}$ and ${\rm ^2P^o_{3/2}-{^2P}^o_{1/2}}$ transitions in Ramsbottom et al. (see Fig.~\ref{S2_ups}). Due to these gross discrepancies, this latter dataset is excluded from further consideration.


\subsection{\ion{S}{3}}

As indicated in Table~\ref{targets}, our target expansion includes the lowest ten $LS$ terms and 31 correlation configurations with $n\leq 4$ orbitals. Previous $R$-matrix work has been carried in $LS$-coupling by \citet{gal95} (15-term target) and \citet{hud12} (29-term target), whereby the IC collision strengths are obtained by a method of algebraic recoupling of the reactance matrices that allows for target fine-structure splittings. The general agreement of the present effective collision strengths with these two datasets is $\sim 25\%$.


\subsection{\ion{Ar}{3}}
\label{ar3}

The target model for this system involves 10 $LS$ terms and correlation configurations with $n\le 4$ orbitals, including an open 2p sub-shell (see Table~\ref{targets}). Agreement with the previous computations by \citet{gal95} and \citet{mun09} is $\sim 20$\% (see Table~\ref{ups}), but a larger discrepancy of a factor of two stands out for the ${3s^23p^4\ ^1D_2}-{^1S_0}$ quadrupole transition due to correlation effects arising from the open 2p sub-shell in the present target model.


\section{$A$-values}
\label{aval}

The forbidden line ratios that are studied in the present work are listed in Table~\ref{lineratios}, and correspond to those employed by observers to derive electron temperatures for the Galactic \ion{H}{2} regions under consideration. The radiative transition probabilities ($A$-values) that we have adopted to determine the line emissivities have been obtained from different compilations---mainly from \citet{bad06}, Appendix~A of \citet{sta12}, and the MCHF/MCDHF (Version~2) database\footnote{\tt http://nlte.nist.gov/MCHF/} \citep{tac01, iri05}---in order to evaluate temperature sensitivity. It is generally found that, for the characteristic  electron densities ($N_e\lesssim 2\times 10^4$~cm$^{-3}$) and temperatures ($T_e\lesssim 15000$~K) of our \ion{H}{2} sources, the line ratios are practically insensitive to the $A$-value choice, i.e. between \citet{bad06} and \citet{tac01} say, except for [\ion{S}{2}]. For this species, the $({\lambda}6716+{\lambda}6731)/({\lambda}4069+{\lambda}4076)$ line ratio, even at the lower densities ($N_e\sim 10^2$~cm$^{-3}$), shows a noticeable $A$-value dependence, and as previously discussed by \citet{tay10}, an undesirable wide scatter is encountered among the published radiative data. We have therefore been encouraged to compute with the atomic structure code {\sc autostructure} new $A$-values for this system with extensive configuration interaction, and in particular, taking into account the Breit correction to the magnetic dipole operator that has been shown by \citet{men82} to be important for ions such as \ion{S}{2} with a ${\rm 3p}^3$ ground configuration.

In Table~\ref{s2aval} we tabulate $A$-values for [\ion{S}{2}] from several theoretical datasets where the aforementioned wide scatter is confirmed. In the present context, the $A$-value ratios
\begin{equation}
R_1=\frac{3}{2}\times\frac{A(^2D^o_{5/2}-{^4S}^o_{3/2})}{A(^2D^o_{3/2}-{^4S}^o_{3/2})}
\end{equation}
and
\begin{equation}
R_2=\frac{A(^2P^o_{3/2}-{^4S}^o_{3/2})+A(^2P^o_{1/2}-{^4S}^o_{3/2})}{A(^2D^o_{5/2}-{^4S}^o_{3/2})+A(^2D^o_{3/2}-{^4S}^o_{3/2})}
\end{equation}
are of interest. As pointed out by \citet{men82}, $R_1$ tends to the ratio of the line intensities
\begin{equation}
R_1\rightarrow \frac{I(^2D^o_{5/2}-{^4S}^o_{3/2})}{I(^2D^o_{3/2}-{^4S}^o_{3/2})}
\end{equation}
as $N_e\rightarrow\infty$ which gives rise to a useful benchmark with observations of dense nebulae. In this respect, \citet{zha05} report a ratio of $R_1=0.443$ for the dense ($N_e=4.7\times 10^4$~cm$^{-3}$) planetary nebula NGC\ 7027, which is only matched accurately (better than $10\%$) in Table~\ref{s2aval} by \citet{men82}, \citet{fri99}, \citet{iri05} (with adjusted level energies), and the present work. $R_2$ is associated to the [\ion{S}{2}] temperature diagnostic, and the agreement of the present value with those by \citet{men82} and \citet{fri99} is within 1\% while only $\sim 10\%$ with \citet{iri05}. The poorer accord with the latter may be due to their exclusion of the relativistic correction to the magnetic dipole operator. In the light of this outcome, the radiative data by \citet{kee93}, \citet{iri05} (with {\em ab-initio} level energies), and \citet{tay10} for [\ion{S}{2}] will not be further discussed.


\section{Temperature diagnostics}
\label{h2temp}

In order to study the impact of the atomic data on nebular temperature diagnostics, we have selected the same group of spectra of  Galactic \ion{H}{2} regions considered by \citet{nic12}, namely
\begin{itemize}
  \item The Orion nebula observed by \citet{est04}
  \item NGC\,3576 observed by \citet{gar04}
  \item S311 observed by \citet{gar05}
  \item M16, M20, and NGC\,3603 observed by \citet{gar06}
  \item HH\,202 in both nebular and shock regions as observed by \citet{mes09}.
\end{itemize}
We have not included here the spectra of extragalactic \ion{H}{2} regions also taken into account by \citet{nic12} as some of the lines of our comprehensive set of temperature diagnostics are not reported. Also, we have not extended our study to planetary nebulae as the CEL fluxes of at least [\ion{N}{2}] and [\ion{O}{2}] in many of these objects are usually contaminated with recombination-line contributions. Although these fluxes are usually corrected with the widely used formula of \citet{liu01}, we do not have a reliable measure of its accuracy. Such flux corrections in the present set of \ion{H}{2} regions, as reported in the observational papers, are found to be less than $\sim 5\%$.

In Table~\ref{temp}, we tabulate for the different \ion{H}{2} regions (in increasing density order) the electron temperatures derived with our selected datasets of effective collision strengths, which allow for each diagnostic the estimate of an average electron temperature $\langle T^{\rm th}_e\rangle$ for comparison with the quoted value, $T^{\rm ob}_e$, in the observational papers. It may be appreciated that the agreement between $\langle T^{\rm th}_e\rangle$ and $T^{\rm ob}_e$ for most diagnostics is better than 10\% except for [\ion{S}{2}] in S311, M20, M16, HH\,202 (shock region), and NGC\,3603 where differences of 20--25\% are found and of $\sim 12\%$ for [\ion{O}{2}] in M20 and Orion. Such discrepancies in our opinion are caused by the use of poor atomic data in previous analyses, i.e. either radiative, collisional or both.

It must be pointed out that the scatter of $\langle T^{\rm th}_e\rangle$ in Table~\ref{temp} is a direct consequence of the inherent statistical uncertainties of resonance phenomena in electron--ion collisional processes, where small variations in quantum mechanical models can give rise to large rate discrepancies. By comparing the magnitudes of the error margins of $\langle T^{\rm th}_e\rangle$ and $T^{\rm ob}_e$ in Table~\ref{temp}, we are inclined to believe that the errors due to the atomic data have not always been taken into account in estimates of $\Delta T^{\rm ob}_e$, and therefore, a more reliable temperature uncertainty would perhaps be $\Delta T^{\rm th}_e + \Delta T^{\rm ob}_e$. In this light, it is then plausible to derive for certain sources, namely M20, S311, and NGC\,3576, an average temperature with a standard deviation comparable to the specified error bars; for other objects, e.g. NGC\,3603, there seems to be at least two well-defined plasma regions associated with ionic excitation.

The noticeable dependence of the [\ion{S}{2}] electron temperature on the $A$-values, specially at the higher densities, is further illustrated in Table~\ref{Aval_temp} where the electron temperatures have been obtained with our effective collision strengths but different radiative datasets. Differences between the present temperatures with those obtained with the $A$-values of \citet{men82} and \citet{fri99} are not larger than 4\% while those derived with the $A$-values by \citet{iri05} are lower by as much as 15\% at the higher densities.


\section{$\kappa$-distribution effects}
\label{kappaeffects}

\begin{figure}
   \rotatebox{0}{\resizebox{\hsize}{\hsize}{\plotone{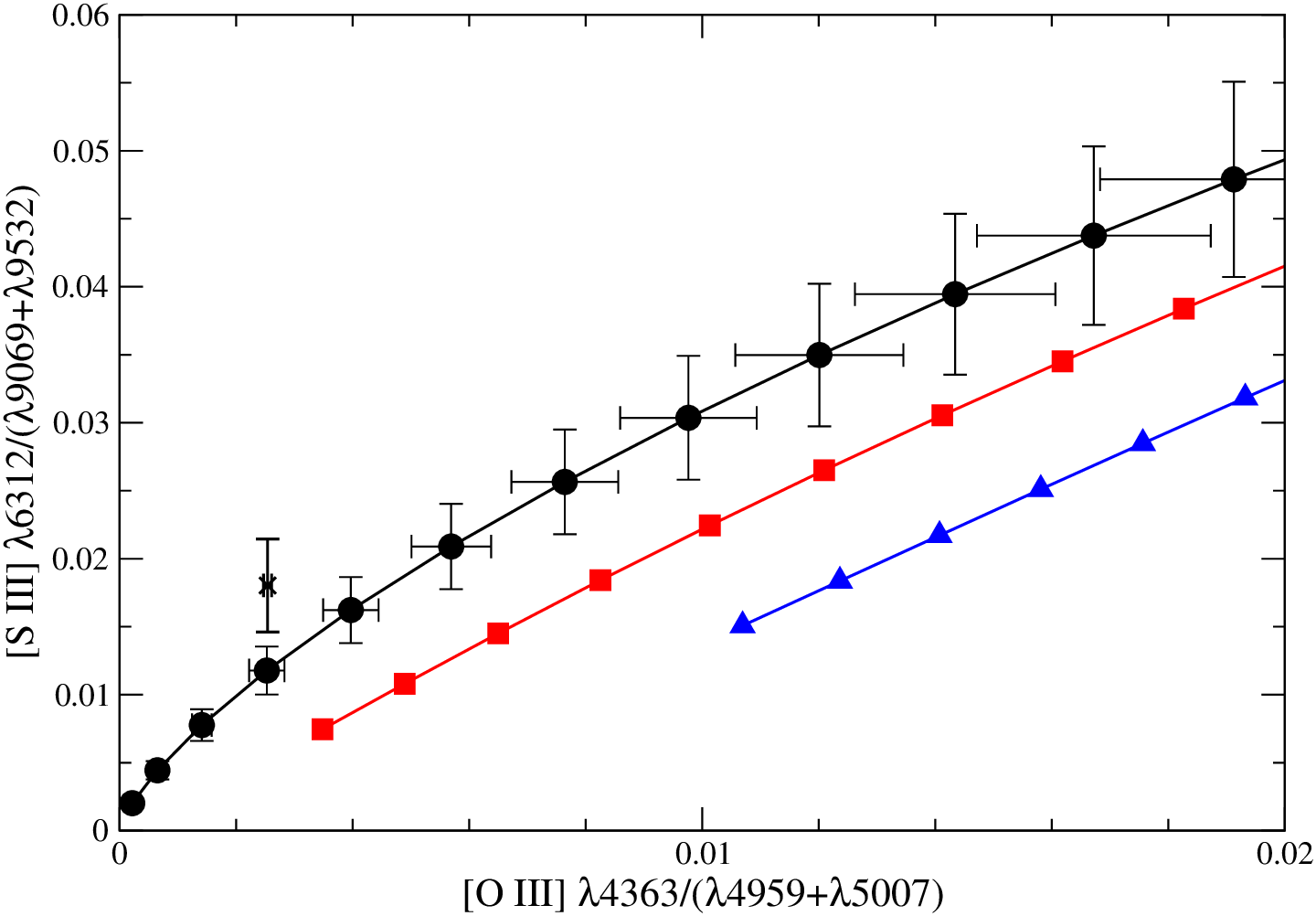}}}
  \caption{Line-ratio map for [\ion{O}{3}] ${\lambda}4363/({\lambda}4959+{\lambda}5007)$ {\em vs}. [\ion{S}{3}] ${\lambda}6312/({\lambda}9069+{\lambda}9352)$ at $N_e=8.9\times 10^3$~cm$^{-3}$. Cross: observed ratio in the Orion nebula \citep{est04}. Black circles: MB distribution. Red squares: $\kappa$ distribution with $\kappa=10$. Blue triangles: $\kappa$ distribution with $\kappa=5$. The points on the curves represent temperature values starting at 5000~K and increasing in steps of 1000~K. \label{orion1}
  }
\end{figure}

\begin{figure}
   \rotatebox{0}{\resizebox{\hsize}{\hsize}{\plotone{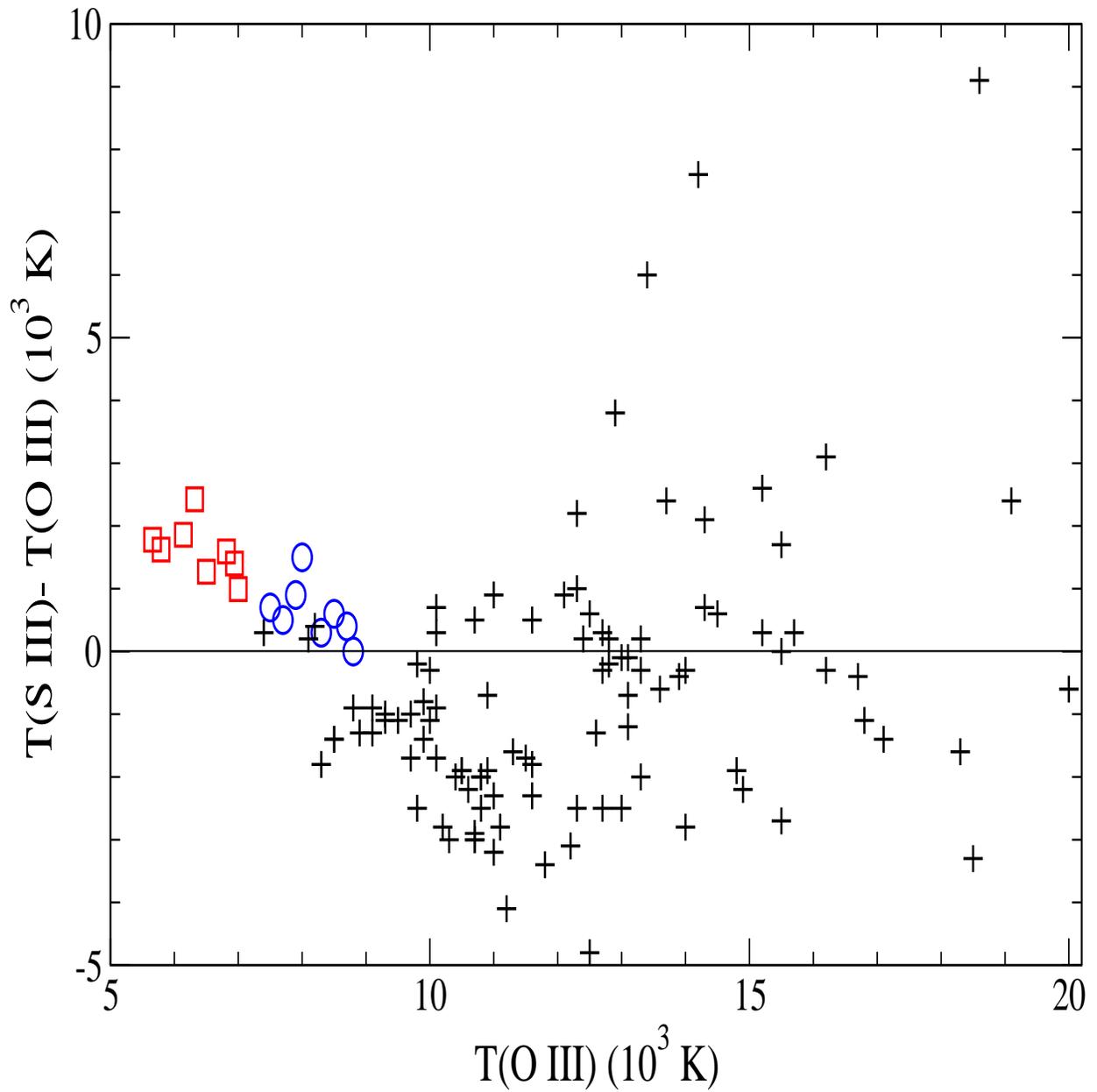}}}
  \caption{Temperature difference between $T$(\ion{S}{3}) and $T$(\ion{O}{3}). Blue circles: present sample of \ion{H}{2} regions assuming a MB distribution. Red squares: present sample assuming a $\kappa$ distribution with $\kappa=20$. Crosses: sample by \citet{bin12}.\label{tdif}
  }

\end{figure}

It may be noted in Table~\ref{temp} that the [\ion{O}{3}] temperature in most objects is distinctively lower than other diagnostics, in particular $T_e($\ion{O}{3}$) < T_e($\ion{S}{3}). Such standing differences have been previously reported by, for instance, \citet{bin12} in an extensive sampling of Galatic and extragalactic \ion{H}{2} regions, but in their case it is the converse: $T_e($\ion{O}{3}$) > T_e($\ion{S}{3}) mostly when gas metallicities are greater than 0.2 solar. In this context, we plot in Fig.~\ref{orion1} the line-ratio map for [\ion{O}{3}] ${\lambda}4363/({\lambda}4959+{\lambda}5007)$ {\em vs}. [\ion{S}{3}] ${\lambda}6312/({\lambda}9069+{\lambda}9352)$ at $N_e=8.9\times 10^3$~cm$^{-3}$. It is seen that, in spite of the inaccuracies in both observation and atomic data, the observed [\ion{O}{3}] and [\ion{S}{3}] line ratios in the Orion nebula \citep{est04} neither conduce to a common MB $T_e$ nor to a common $T_\kappa$ when different values of $\kappa$ are considered. This trend prevails in all the \ion{H}{2} regions of our sample as further illustrated in Fig.~\ref{tdif}: for MB, $0\lesssim T_e($\ion{S}{3}$)-T_e($\ion{O}{3}$)\lesssim 1.5\times 10^3$~K while for $\kappa=20$ both temperatures are significantly reduced but now $10^3\lesssim T_\kappa($\ion{S}{3}$)-T_\kappa($\ion{O}{3}$)\lesssim 2.4\times 10^3$~K; i.e. the temperature differences are significantly increased ($\sim 10^3$~K) by the $\kappa$~distribution. In Fig.~\ref{tdif} we also include the temperatures determined in the larger sample of \citet{bin12}, where the scatter is much larger ($-5\times 10^3\lesssim T_e($\ion{S}{3}$)-T_e($\ion{O}{3}$)\lesssim 10^4$~K) and, in contrast to the present dataset, most of their objects have $T_e($\ion{O}{3}$)> 10^4$~K. It must be noted that in Fig.~\ref{tdif} we have plotted the $\Delta T{-}T$ relation rather than the usual $T{-}T$ since we would like to emphasize several points: $\Delta T$ is found to take positive and negative values; the $T$ reduction by the $\kappa$~distribution does not necessarily lead to a diminished $\Delta T$ as desired; and unlike previous reports, we have avoided displaying the apparent $\kappa$-distribution $T$ as it is density dependent and, thus, arguably reliable for a large nebula sample such as that of \citet{bin12} for which we are unaware of the density range.

A similar situation is found in the [\ion{O}{3}] {\em vs.} [\ion{Ar}{3}] line-ratio map of our sample which implies that, in the high-excitation region, the diagnostic temperature differences do not seem to be caused by non-Maxwellian distributions. Evidently, this conclusion cannot be extended to the sample by \citet{bin12} where the $\kappa$~distribution may indeed contribute to reduce some of the [\ion{O}{3}] and [\ion{S}{3}] temperature disparities, at least for the large population for which $T_e($\ion{S}{3}$)-T_e($\ion{O}{3}$)<0$~K (see Fig.~\ref{tdif}).

\begin{figure}
   \rotatebox{0}{\resizebox{\hsize}{\hsize}{\plotone{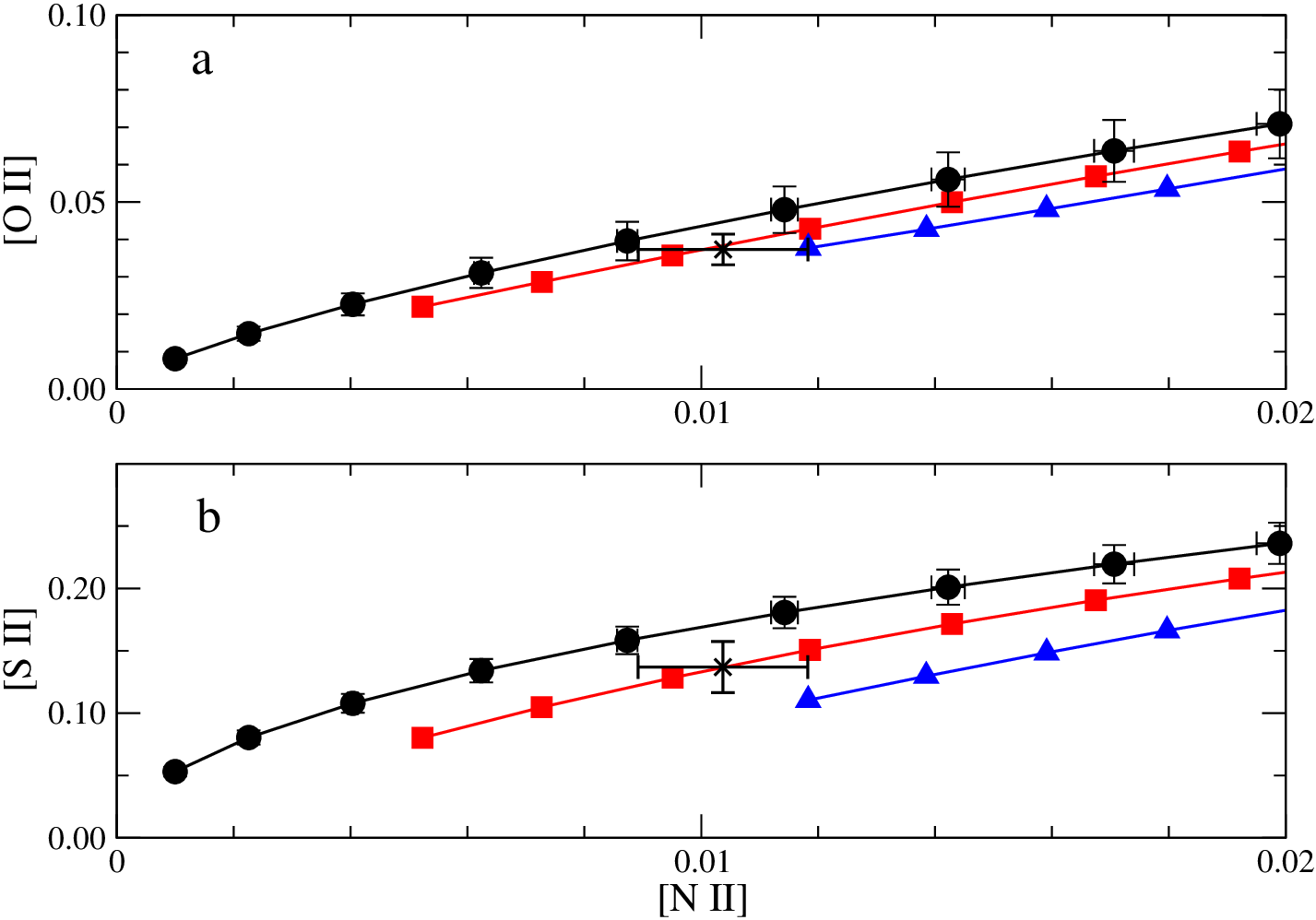}}}
  \caption{Line-ratio maps for (a) [\ion{N}{2}] ${\lambda}5755/({\lambda}6548+{\lambda}6583)$ {\em vs}. [\ion{O}{2}] $({\lambda}7320+{\lambda}7330)/({\lambda}3726+{\lambda}3729)$ and (b) [\ion{N}{2}] ${\lambda}5755/({\lambda}6548+{\lambda}6583)$ {\em vs}. [\ion{S}{2}] $({\lambda}4069+{\lambda}4076)/({\lambda}6716+{\lambda}6731)$ at $N_e=2.9\times 10^3$~cm$^{-3}$. Crosses: observed ratios in HH\,202 (nebular component) by \citet{mes09}. Black circles: MB distribution. Red squares: $\kappa$ distribution with $\kappa=10$. Blue triangles: $\kappa$ distribution with $\kappa=5$. The points on the curves represent temperature values starting at 5000~K and increasing in steps of 1000~K. \label{HH202}
  }
\end{figure}

For the low-excitation region, the situation is somewhat different as the $\kappa$~distribution can lead in some circumstances to a lower common temperature for the [\ion{N}{2}], [\ion{O}{2}], and [\ion{S}{2}] diagnostics. This is the case, as shown in Fig.~\ref{HH202}, of the nebular component of HH\,202 where the three different MB temperatures $T_e({\rm N~\textsc{ii}})=9700$~K, $T_e({\rm O~\textsc{ii}})=8800$~K, and $T_e({\rm S~\textsc{ii}})=8300$~K
can be reduced to a single, lower temperature of $T_\kappa\approx 7400$~K with $\kappa=10$. As a result, the lower temperature leads to abundance increases for \ion{N}{2}, \ion{O}{2}, and \ion{S}{2} of factors of 1.79, 2.21, and 1.32, respectively. It must be pointed out that the reliability of these findings is highly limited by the confidence levels of both the observations and atomic data; for instance, it may be seen in Fig.~\ref{HH202} that the error bar of the [\ion{N}{2}] line ratio is fairly large ($\sim 19\%$) thus hampering the choice of an accurate $\kappa$ value or even a reliable departure from the MB distribution.

\begin{figure}
   \rotatebox{0}{\resizebox{\hsize}{\hsize}{\plotone{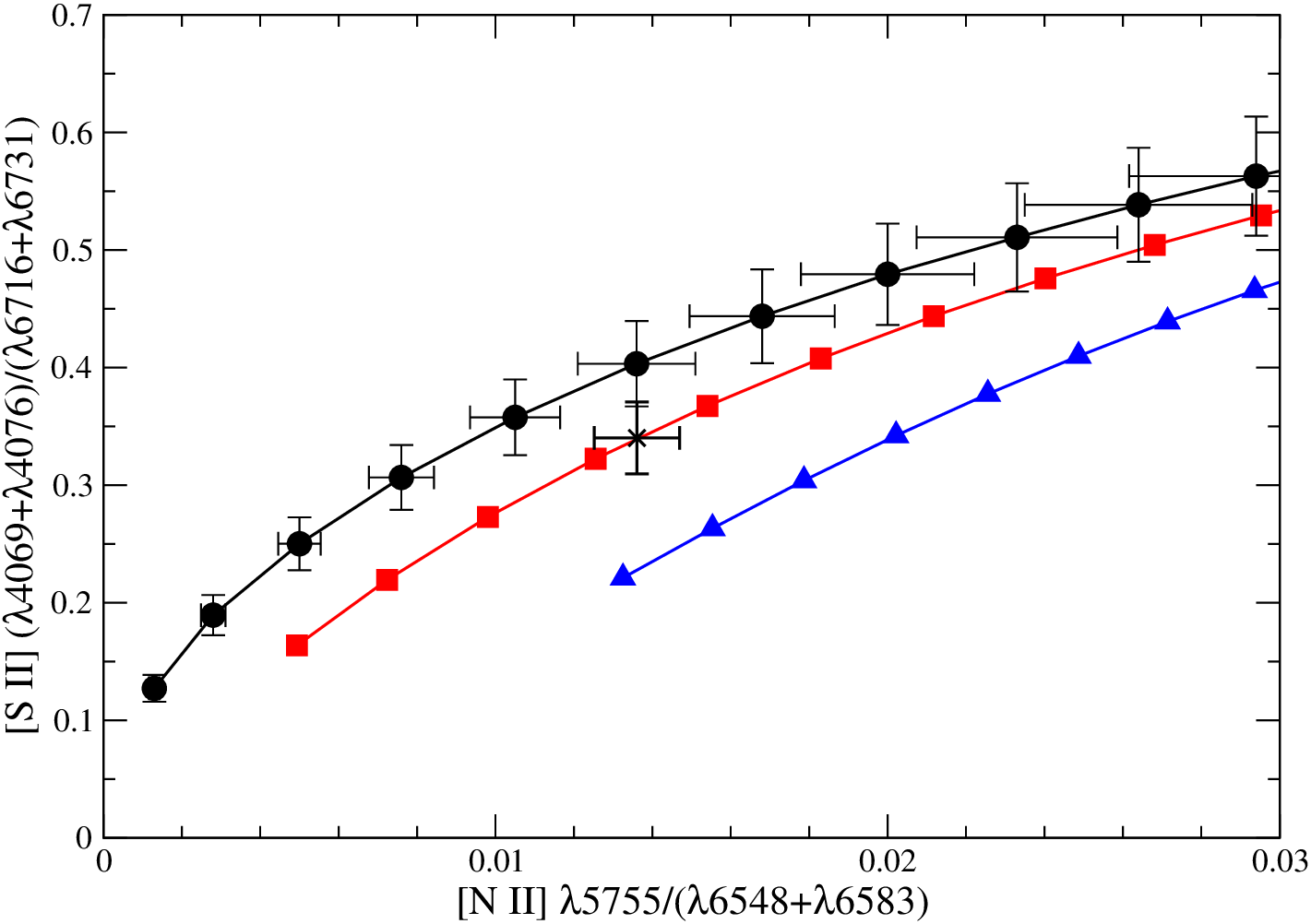}}}
  \caption{Line-ratio map for [\ion{N}{2}] ${\lambda}5755/({\lambda}6548+{\lambda}6583)$ {\em vs}. [\ion{S}{2}] $({\lambda}4069+{\lambda}4076)/({\lambda}6716+{\lambda}6731)$ at $N_e=8.9\times 10^3$~cm$^{-3}$. Cross: observed ratio in the Orion nebula \citep{est04}. Black circles: MB distribution. Red squares: $\kappa$ distribution with $\kappa=12$. Blue triangles: $\kappa$ distribution with $\kappa=5$. The points on the curves represent temperature values starting at 5000~K and increasing in steps of 1000~K. \label{orion2}
  }
\end{figure}

In the Orion nebula (see Fig.~\ref{orion2}), the two MB temperatures of $T_e({\rm N~\textsc{ii}})=10000$~K and $T_e({\rm S~\textsc{ii}})=8700$~K can also be reduced to a common $\kappa$~temperature of $T_\kappa\approx 8400$~K with $\kappa=12$, the abundance enhancements for \ion{N}{2} and \ion{S}{2} in this case being smaller: 32\% and 8\%, respectively; however, this procedure cannot be extended to the low $T_e({\rm O~\textsc{ii}})=7900$~K which could perhaps be due to errors in our atomic data since the temperatures obtained with other collisionally datasets are significantly higher (see Table~\ref{temp}). Furthermore, in other sources, e.g. M16, there is no conclusive evidence that the $\kappa$~distribution would lead to a common lower temperature in the low-excitation region.


\section{Summary and conclusions}
\label{summary}

It has become clear in the work of \citet{nic13} and \citet{dop13} on non-Maxwellian electron distributions in nebular plasmas  that reliable effective collision strengths and $A$-values are essential before any recommendations can be drawn. We have therefore been encouraged to make an attempt at quantifying the impact of such atomic data on the electron temperatures of a sample of Galactic \ion{H}{2} regions for which accurate spectra are available, and to search for evidence to substantiate departures from the MB framework. This initiative has implied extensive computations of new collisional datasets for the ionic systems that give rise in nebulae to CEL plasma diagnostics in order to derive both MB and $\kappa$-averaged rate coefficients with a comparable degree of accuracy. These energy-tabulated collision strengths are currently available for download from the AtomPy\footnote{\tt http://bit.ly/K5oAfD} data curation service.

By extensive comparisons with other datasets of effective collision strengths computed in the past two decades, we have found that their statistical consistency is around the $\sim 20{-}30$\% level although larger discrepancies (factor of 2, say) may be detected for the odd transition. This inherent dispersion, which in our opinion would be hard to reduce in practice, is due in forbidden transitions to the strong sensitivity of the resonance contribution to the scattering numerical approach, target atomic model, energy mesh, and small interactions such as relativistic corrections. However, with the exception of the difficult [\ion{S}{2}] system, the theoretical temperature dispersion in most diagnostics has been found to be not larger than 10\%, and the agreement with the temperatures estimated in the observational papers is within this similar satisfactory range. Therefore, there is room for optimism. As shown in the present analysis, $T_e$(\ion{S}{2}) has been proven to be strongly dependent on both the radiative and collisional datasets, and in this context, only the $A$-values by \citet{men82}, \citet{fri99}, and the present comply with all the stringent specifications required. On the other hand, the collisional dataset for this ion still needs more refinement before a firm ranking can be put forward on its accuracy.

In the present study, which has relied on high-quality sets of astronomically observed line intensities for \ion{H}{2} regions with fairly low electron temperatures ($T_e\lesssim 1.5\times 10^4$~K), it has been shown that the error margins due to the atomic data are not always taken into account in observational work when quoting electron temperature uncertainties. When they are included, rather than trying to differentiate plasma regions, an average temperature may be defined in some sources with a standard deviation comparable to the error margins of each temperature diagnostic. If multi-region plasmas are indeed in evidence, the observed diagnostic temperature variation in our sample might be resolved by $\kappa$~distributions only in the low-excitation region, thus leading to a single lower temperature for diagnostics such as [\ion{N}{2}], [\ion{O}{2}], and [\ion{S}{2}] with the consequent abundance enhancements. However, even in these cases the differences between observed line ratios and the predictions of the MB distribution differ by less than $2\sigma$ if proper uncertainties are taken into account. In the high-excitation region, on the other hand, diagnostic temperature differences in [\ion{O}{3}], [\ion{S}{3}], and [\ion{Ar}{3}] do not seem to be reconciled by invoking $\kappa$~distributions, which then leaves the problem open to other unaccounted physical processes. It must be emphasized that these findings are not the rule as indicated by the extensive sample by \citet{bin12} of \ion{H}{2} regions with mostly higher temperature ($T_e> 10^4$~K); thus further detailed work would be required before $\kappa$ distributions can be firmly established.

Regarding the effects of the $\kappa$~distribution on the collisional rate coefficients, we have confirmed the reliability of their neglect for de-excitation, and an accurate (better than 10\%) analytic expression has been derived to readily estimate $\kappa$-averaged rates for excitation from temperature tabulations of MB effective collision strengths. The measure that regulates level-excitation departures from MB is $T/T_{\rm ex}\ll 1$; hence, for IR transitions within the ionic ground terms with very low excitation energies ($T_{\rm ex} < 10^3$~K), the MB and $\kappa$-distribution electron impact excitation rates are essentially identical at nebular temperatures. IR abundance discrepancies with respect to the optical would then be a further indication of the relevance of $\kappa$~distributions. However, such discrepancies have not been reported; for instance, all planetary nebulae observed with {\it ISO} show consistent abundances as estimated from both IR and optical CEL \citep{wes05}, which then suggests that the MB distribution dominates.

In the present work it has been demonstrated that error-margin capping in the collisional data relevant to nebular diagnostic implies---apart from detailed comparisons with previous datasets and in some cases recalculations---comprehensive benchmarks with spectral models and observations, and this is a task we intend to pursue in the ongoing discussion concerning the CEL--RL abundance discrepancies.


\acknowledgements

We would like to thank Christophe Morisset, Instituto de Astronom\'ia, UNAM, Mexico, for sending us the temperature diagnostics for the \ion{H}{2} region sample by \citet{bin12}. We are also indebted to Grazyna Stasi{\'n}ska, Observatoire de Paris, France, for useful discussions regarding several aspects of the paper. 


\bibliographystyle{apj}

\newpage

\begin{deluxetable}{ll}
\tablewidth{0pt}
\tablecaption{Forbidden-line diagnostics \label{lineratios}}
\tablecolumns{2}
\tablehead{\colhead{System} & \colhead{Line ratio}}
\startdata
[N~{\sc ii}]   & $({\lambda}6548+{\lambda}6583)/{\lambda}5755$ \\

[O~{\sc ii}]   & $({\lambda}3726+{\lambda}3729)/({\lambda}7320+{\lambda}7330)$ \\

[O~{\sc iii}]  & $({\lambda}4959+{\lambda}5007)/{\lambda}4363$ \\

[S~{\sc ii}]   & $({\lambda}6716+{\lambda}6731)/({\lambda}4069+{\lambda}4076)$ \\

[S~{\sc iii}]  & $({\lambda}9069+{\lambda}9532)/{\lambda}6312$ \\

[Ar~{\sc iii}] & $({\lambda}7136+{\lambda}7751)/{\lambda}5192$ \\
\enddata
\end{deluxetable}


\begin{deluxetable}{lll}
\tabletypesize{\scriptsize}
\tablewidth{0pt}
\tablecaption{Target representations \label{targets}}
\tablecolumns{3}
\tablehead{\colhead{Ion} & \colhead{Target terms} & \colhead{Correlation configurations}}
\startdata
N~{\sc ii} & 2s$^2$2p$^2$\ $^3$P, $^1$D, $^1$S & 2p$^4$, 2s$^2$2p3p, 2s$^2$2p3d, 2s2p$^2$3s, \\
           & 2s2p$^3$\ $^5$S$^{\rm o}$, $^3$D$^{\rm o}$, $^3$P$^{\rm o}$, $^1$D$^{\rm o}$, $^3$S$^{\rm o}$
           & 2s2p$^2$3p, 2s2p$^2$3d, 2s$^2$3s$^2$, 2s$^2$3p$^2$, \\
           & 2s$^2$2p3s\ $^1$P$^{\rm o}$, $^3$P$^{\rm o}$ & 2s$^2$3d$^2$, 2p$^3$3p, 2p$^3$3d \\
\hline
O~{\sc ii} & 2s$^2$2p$^3$\ $^4$S$^{\rm o}$, $^2$D$^{\rm o}$, $^2$P$^{\rm o}$ & 2s$^3$2p$^2$3d, 2s2p$^3$3s, 2s2p$^3$3p, 2s2p$^3$3d, \\
           & 2s2p$^4$\ $^4$P, $^2$D, $^2$S, $^2$P &  2s$^2$2p3s$^2$, 2s$^2$2p3p$^2$, 2s$^2$2p3d$^2$ \\
           & 2s$^2$2p$^2$3s\ $^4$P, $^2$P, $^2$D, $^2$S &  \\
           & 2s$^2$2p$^2$3p\ $^2$S$^{\rm o}$, $^4$D$^{\rm o}$, $^4$P$^{\rm o}$, $^2$D$^{\rm o}$, $^4$S$^{\rm o}$, & \\
           & \ \ \ \ \ \ \ \ \ \ \ \ \ \ $^2$P$^{\rm o}$, $^2$F$^{\rm o}$, $^2$D$^{\rm o}$ & \\
\hline
O~{\sc iii} & 2s$^2$2p$^2$\ $^3$P, $^1$D, $^1$S & 2p$^4$, 2s$^2$2p3s, 2s$^2$2p3p, 2s$^2$2p3d, 2s2p$^2$3s, \\
            & 2s2p$^3$\ $^5$S$^{\rm o}$, $^3$D$^{\rm o}$, $^3$P$^{\rm o}$, $^1$D$^{\rm o}, ^3$S$^{\rm o}$, $^1$P$^{\rm o}$
            &  2s2p$^2$3p, 2s2p$^2$3d, 2s$^2$3s$^2$, 2s$^2$3p$^2$, \\
            & & 2s$^2$3d$^2$, 2s$^2$3s3p, 2s$^2$3s3d, 2p$^3$3p, 2p$^3$3d\\
\hline
S~{\sc ii} & 3s$^2$3p$^3$\ $^4$S$^{\rm o}$, $^2$D$^{\rm o}$, $^2$P$^{\rm o}$ & 2p$^6$3p$^5$, 3s3p$^2$3d4s, 3s$^2$3p3d4s4p, 3s$^2$3p4s$^2$, \\
           & 3s3p$^4$\ $^4$P, $^2$D, $^2$S & 3s3p$^3$4p, 3s3p$^2$4p$^2$, 3s3p$^2$3d4p, 2p$^5$3s$^2$3p$^4$, \\
           & 3s$^2$3p$^2$3d\ $^2$P, $^4$F, $^4$D, $^2$F, $^4$P  & 2p$^5$3s$^2$3p$^3$3d, 2p$^5$3s$^2$3p$^3$4s, 2p$^5$3s$^2$3p$^3$4p \\
           & 3s$^2$3p$^2$4s\ $^4$P, $^2$P    & \\
           & 3s$^2$3p$^2$4p\ $^4$P$^{\rm o}$, $^2$D$^{\rm o}$, $^4$S$^{\rm o}$, $^2$P$^{\rm o}$ & \\
\hline
S~{\sc iii} & 3s$^2$3p$^2$\ $^3$P, $^1$D, $^1$S & 3p$^4$, 3s$^2$3p4s, 3s$^2$3p4p, 3s$^2$3d$^2$, 3s$^2$3d4s,  \\
            & 3s3p$^3$\ $^5$S$^{\rm o}$, $^3$D$^{\rm o}$, $^1$P$^{\rm o}$, $^3$P$^{\rm o}$, $^3$S$^{\rm o}$
            & 3s$^2$3d4p, 3s$^2$3d4d, 3s$^2$4s$^2$, 3s$^2$4s4p, 3s$^2$4s4d,  \\
            & 3s$^2$3p3d\ $^1$D$^{\rm o}$, $^3$F$^{\rm o}$
            & 3s$^2$4p$^2$, 3s$^2$4p4d, 3s$^2$4d$^2$, 3s3p$^2$3d, 3s3p$^2$4s, \\
            & & 3s3p$^2$4p, 3s3p$^2$4d, 3s3p3d$^2$, 3s3p3d4s, 3s3p3d4p,   \\
            & & 3s3p3d4d, 3s3p4s$^2$, 3s3p4s4p, 3s3p4s4d, 3s3p4p$^2$, \\
            & & 3s3p4p4d, 3s3p4d$^2$, 3p$^3$3d, 3p$^3$4s, 3p$^3$4p, 3p$^3$4d \\
\hline
Ar~{\sc iii} & 2p$^6$3s$^2$3p$^4$\ $^3$P, $^1$D, $^1$S & 2p$^6$3p$^6$,
2p$^6$3s3p$^4$3d, 2p$^6$3s$^2$3p$^2$3d$^2$, 2p$^6$3p$^4$3d$^2$\\
             & 2p$^6$3s3p$^5$\ $^3$P, $^1$P & 2p$^6$3s$^2$3p$^3$4p,
2p$^6$2p$^5$4s, 2p$^6$3s$^2$3p$^3$4p,2p$^6$3s$^2$3p$^3$4f, \\
             & 2p$^6$3s3p$^3$3d\ $^5$D$^{\rm o}$, $^3$D$^{\rm o}$, $^3$F$^{\rm o}$, $^1$F$^{\rm o}$, $^3$S$^{\rm o}$ & 2p$^5$3s$^2$3p$^4$3d,
2p$^5$3s$^2$3p$^3$3d4s, 2p$^5$3s$^2$3p$^3$3d4d \\
\enddata
\end{deluxetable}


\begin{deluxetable}{lll}
\tablewidth{0pt}
\tablecaption{Effective collision strengths for leading transitions at $10^4$~K \label{ups}}
\tablecolumns{2}
\tablehead{\colhead{Diagnostic} & \colhead{Leading transitions} & \colhead{$\Upsilon^{\rm MB}_{ji}$}}
\startdata
[N~{\sc ii}]   & ${\rm 2s^22p^2\ ^1D_2}-{\rm ^3P_0}$ & 0.284,$^a$ 0.293,$^b$ 0.286$^c$ \\

               & ${\rm 2s^22p^2\ ^1S_0}-{\rm ^3P_0}$ & 0.0327,$^a$ 0.0326,$^b$ 0.0333$^c$ \\

[O~{\sc ii}]   & ${\rm 2s^22p^3\ ^2D^o_{5/2}}-{\rm ^4S^o_{3/2}}$ & 0.973,$^a$ 0.883,$^d$ 0.803,$^e$ 0.834$^f$ \\

               & ${\rm 2s^22p^3\ ^2P^o_{3/2}}-{\rm ^4S^o_{3/2}}$ & 0.356,$^a$ 0.313,$^d$ 0.283,$^e$ 0.256$^f$ \\

[O~{\sc iii}]  & ${\rm 2s^22p^2\ ^1D_2}-{\rm ^3P_0}$ & 0.220,$^a$ 0.254,$^b$ 0.243,$^g$ 0.269$^h$ \\

               & ${\rm 2s^22p^2\ ^1S_0}-{\rm ^3P_0}$ & 0.0283,$^a$ 0.0325,$^b$ 0.0321,$^g$ 0.0407$^h$ \\

[S~{\sc ii}]   & $3s^23p^3\ {^2D^o_{5/2}}- {^4S^o_{3/2}}$ & 3.99,$^a$ 4.66,$^i$ 3.83$^j$ \\

               & ${\rm 3s^23p^3\ ^2P^o_{3/2}}-{\rm ^4S^o_{3/2}}$ & 1.53,$^a$ 1.38,$^i$ 1.42$^j$ \\

[S~{\sc iii}]  & ${\rm 3s^23p^2\ ^1D_2}-{\rm ^3P_0}$ &  0.868,$^a$ 0.879,$^k$ 0.729$^l$ \\

               & $3s^23p^2\ {^1S_0}-{^3P_0}$ &  0.159,$^a$ 0.122,$^k$ 0.125$^l$ \\
               
[Ar~{\sc iii}] & $3s^23p^4\ {^1D_2}-{^3P_2}$ &  3.23,$^a$ 2.66,$^k$ 2.94$^m$ \\

               & $3s^23p^4\ {^1S_0}-{^3P_2}$ &  0.422,$^a$ 0.463,$^k$ 0.354$^m$ \\
\enddata
\tablenotetext{a}{Present work}
\tablenotetext{b}{\citet{len94}}
\tablenotetext{c}{\citet{tay11}}
\tablenotetext{d}{\citet{pra06}}
\tablenotetext{e}{\citet{tay07}}
\tablenotetext{f}{\citet{kis09}}
\tablenotetext{g}{\citet{agg99}}
\tablenotetext{h}{\citet{pal12}}
\tablenotetext{i}{\citet{cai93}}
\tablenotetext{j}{\citet{tay10}}
\tablenotetext{k}{\citet{gal95}}
\tablenotetext{l}{\citet{hud12}}
\tablenotetext{m}{\citet{mun09}}
\end{deluxetable}


\begin{deluxetable}{llllllll}
\tabletypesize{\scriptsize}
\tablewidth{0pt}
\tablecaption{Theoretical $A$-values (s$^{-1}$) for transitions within the \ion{S}{2} $3s^23p^3$ ground configuration\label{s2aval}}
\tablecolumns{8}
\tablehead{\colhead{Parameter} & \colhead{Pres$^a$} & \colhead{MZ$^b$} & \colhead{KHO$^c$} & \colhead{FFG$^d$} & \colhead{IFF1$^e$} & \colhead{IFF2$^f$} & \colhead{TZ$^g$}}
\startdata
$A(^2D^o_{3/2}-{^4S}^o_{3/2})$ & 8.95E$-$4 & 8.82E$-$4 & 1.24E$-$3 & 9.12E$-$4 & 7.26E$-$4 & 6.84E$-$4 & 6.32E$-$4 \\
$A(^2D^o_{5/2}-{^4S}^o_{3/2})$ & 2.66E$-$4 & 2.60E$-$4 & 2.85E$-$4 & 2.51E$-$4 & 2.26E$-$4 & 2.02E$-$4 & 2.20E$-$4 \\
$A(^2D^o_{5/2}-{^2D}^o_{3/2})$ & 3.46E$-$7 & 3.35E$-$7 & 9.35E$-$7 & 4.10E$-$7 & 2.43E$-$7 & 2.34E$-$7 & 1.71E$-$7 \\
$A(^2P^o_{1/2}-{^4S}^o_{3/2})$ & 9.24E$-$2 & 9.06E$-$2 & 1.14E$-$1 & 8.96E$-$2 & 7.83E$-$2 & 7.74E$-$2 & 7.64E$-$2 \\
$A(^2P^o_{1/2}-{^2D}^o_{3/2})$ & 1.53E$-$1 & 1.63E$-$1 & 1.53E$-$1 & 1.58E$-$1 & 1.35E$-$1 & 1.43E$-$1 & 1.47E$-$1 \\
$A(^2P^o_{1/2}-{^2D}^o_{5/2})$ & 7.09E$-$2 & 7.79E$-$2 & 6.27E$-$2 & 7.18E$-$2 & 6.35E$-$2 & 6.87E$-$2 & 7.16E$-$2 \\
$A(^2P^o_{3/2}-{^4S}^o_{3/2})$ & 2.29E$-$1 & 2.25E$-$1 & 2.84E$-$1 & 2.28E$-$1 & 1.95E$-$1 & 1.92E$-$1 & 1.90E$-$1 \\
$A(^2P^o_{3/2}-{^2D}^o_{3/2})$ & 1.27E$-$1 & 1.33E$-$1 & 1.40E$-$1 & 1.32E$-$1 & 1.10E$-$1 & 1.15E$-$1 & 1.17E$-$1 \\
$A(^2P^o_{3/2}-{^2D}^o_{5/2})$ & 1.68E$-$1 & 1.79E$-$1 & 1.63E$-$1 & 1.71E$-$1 & 1.47E$-$1 & 1.56E$-$1 & 1.61E$-$1 \\
$A(^2P^o_{3/2}-{^2P}^o_{1/2})$ & 9.13E$-$7 & 1.03E$-$6 & 1.17E$-$6 & 1.63E$-$6 & 2.58E$-$7 & 2.51E$-$7 & 2.43E$-$7 \\
$R_1$                          & 4.46E$-$1 & 4.42E$-$1 & 3.46E$-$1 & 4.13E$-$1 & 4.67E$-$1 & 4.43E$-$1 & 5.21E$-$1 \\
$R_2$                          & 2.77E$+$2 & 2.76E$+$2 & 2.62E$+$2 & 2.73E$+$2 & 2.87E$+$2 & 3.04E$+$2 & 3.13E$+$2 \\
\enddata
\tablecomments{$R_1=\frac{3}{2}\times\frac{A(^2D^o_{5/2}-{^4S}^o_{3/2})}{A(^2D^o_{3/2}-{^4S}^o_{3/2})}$ and $R_2=\frac{A(^2P^o_{3/2}-{^4S}^o_{3/2})+A(^2P^o_{1/2}-{^4S}^o_{3/2})}{A(^2D^o_{5/2}-{^4S}^o_{3/2})+A(^2D^o_{3/2}-{^4S}^o_{3/2})}$}
\tablenotetext{a}{Present work}
\tablenotetext{b}{\citet{men82}}
\tablenotetext{c}{\citet{kee93}}
\tablenotetext{d}{\citet{fri99}}
\tablenotetext{e}{\citet{iri05}, calculated with {\em ab-initio} level energies}
\tablenotetext{f}{\citet{iri05}, calculated with adjusted level energies}
\tablenotetext{g}{\citet{tay10}}
\end{deluxetable}


\begin{deluxetable}{lrllll}
\tabletypesize{\scriptsize}
\tablewidth{0pt}
\tablecaption{Temperature diagnostics in \ion{H}{2} regions \label{temp}}
\tablecolumns{6}
\tablehead{\colhead{H~{\sc ii} region} & \colhead{$N_e$} & \colhead{Diagnostic} & \colhead{$T^{\rm th}_e$}
                            & \colhead{$\langle T^{\rm th}_e\rangle$} & \colhead{$T^{\rm ob}_e$}\\
                            & \colhead{(cm$^{-3}$)} & & \colhead{(K)} & \colhead{(K)} & \colhead{(K)}
}
\startdata
 M20          & 270   & [N~{\sc ii}] & 8500,$^a$ 8600,$^b$ 8400$^c$             & $8500\pm 100$ & $8500\pm 240^n$ \\

              &       & [O~{\sc ii}] & 7900,$^a$ 8100,$^d$ 8300,$^e$ 8600$^f$   & $8200\pm 350$ & $8275\pm 400^n$ \\

              &       & [S~{\sc ii}]  & 8800,$^a$ 10000,$^g$ 9000$^h$           & $9300\pm 600$ & $6950\pm 350^n$ \\

              &       & [O~{\sc iii}]& 7700,$^a$ 7900,$^b$ 7800,$^i$ 7500$^j$   & $7700\pm 200$ & $7800\pm 300^n$ \\

              &       & [S~{\sc iii}] & 8200,$^a$ 9100,$^k$ 8600$^l$            & $8600\pm 450$ & $8300\pm 400^n$ \\

              &       & [Ar~{\sc iii}] & 8900,$^a$ 8700,$^k$ 9800$^m$           & $9100\pm 600$ & $8730\pm 920^n$ \\

S311          & 310   & [N~{\sc ii}] & 9600,$^a$ 9700,$^b$ 9500$^c$             & $9600\pm 100$ & $9500\pm 250^r$ \\

              &       & [O~{\sc ii}] & 9200,$^a$ 9600,$^d$, 9900,$^e$ 10200$^f$ & $9700\pm 450$ & $9800\pm 600^r$ \\

              &       & [S~{\sc ii}] & 9300,$^a$ 10700,$^g$ 9600$^h$            & $9800\pm 700$ & $7200^{+750r}_{-600}$ \\

              &       & [O~{\sc iii}] & 8700,$^a$ 9000,$^b$ 8900,$^i$ 8600$^j$  & $8800\pm 200$ & $9000\pm 200^r$ \\

              &       & [S~{\sc iii}] & 9100,$^a$ 10200,$^k$ 9500$^l$           & $9600\pm 600$ & $9300\pm 350^r$ \\

              &       & [Ar~{\sc iii}] & 8900,$^a$ 8800,$^k$ 9800$^m$           & $9200\pm 600$ & $8800^{+700r}_{-850}$ \\

 M16          & 1100  & [N~{\sc ii}] & 8500,$^a$ 8600,$^b$ 8400$^c$             & $8500\pm 100$ & $8450\pm 270^n$ \\

              &       & [O~{\sc ii}] & 7600,$^a$ 7900,$^d$ 8300,$^e$ 8400$^f$   & $8000\pm 400$ & $8260\pm 400^n$ \\

              &       & [S~{\sc ii}] & 9200,$^a$ 10300,$^g$ 9900$^h$            & $9800\pm 600$ & $7520\pm 430^n$ \\

              &       & [O~{\sc iii}] & 7500,$^a$ 7700,$^b$ 7600,$^i$ 7300$^j$  & $7500\pm 150$ & $7650\pm 250^n$ \\

              &       & [S~{\sc iii}] & 8200,$^a$ 9100,$^k$ 8700$^l$            & $8700\pm 400$ & $8430\pm 450^n$ \\

 NGC\,3576    & 2800  & [N~{\sc ii}] & 8800,$^a$ 8800,$^b$ 8700$^c$             & $8800\pm 100$ & $8500\pm 200^s$ \\

              &       & [O~{\sc ii}] & 7800,$^a$ 8200,$^d$ 8800,$^e$ 8800$^f$   & $8400\pm 500$ & \\

              &       & [S~{\sc ii}] & 7600,$^a$ 8200,$^g$ 7700$^h$             & $7800\pm 300$ & $8400^{+350s}_{-600}$ \\

              &       & [O~{\sc iii}] & 8300,$^a$ 8600,$^b$ 8500,$^i$ 8000$^j$  & $8400\pm 250$ & $8500\pm 50^s$ \\

              &       & [S~{\sc iii}] & 8600,$^a$ 9500,$^k$ 9100$^l$            & $9100\pm 500$ & $9300^{+500s}_{-400}$ \\

              &       & [Ar~{\sc iii}] & 8700,$^a$ 8600,$^k$ 9600$^m$           & $9000\pm 600$ & $8600^{+450s}_{-350}$ \\

 HH\,202 (neb)& 2900  & [N~{\sc ii}] & 9700,$^a$ 9700,$^b$ 9500$^c$             & $9700\pm 100$ & $9610\pm 390^t$ \\

              &       & [O~{\sc ii}] & 8800,$^a$ 7500,$^d$ 7900,$^e$ 7900$^f$   & $8000\pm 600$ & $8790\pm 250^t$ \\

              &       & [S~{\sc ii}] & 8300,$^a$ 9000,$^g$ 9100$^h$             & $8800\pm 400$ & $8010\pm 440^t$ \\

              &       & [O~{\sc iii}] & 7900,$^a$ 8200,$^b$ 8100,$^i$ 7800$^j$  & $8000\pm 200$  & $8180\pm 200^t$ \\

              &       & [S~{\sc iii}] & 8800,$^a$ 9800,$^k$ 9400$^l$            & $9300\pm 500$  &  $8890\pm 270^t$ \\

              &       & [Ar~{\sc iii}] & 8600,$^a$ 7900,$^k$ 8800$^m$           & $8400\pm 500$  & $7920\pm 450^t$ \\

 NGC\,3603    & 5200  & [N~{\sc ii}] & 11200,$^a$ 11200,$^b$ 11000$^c$          & $11100\pm 150$ & $11050\pm 800^n$ \\

              &       & [O~{\sc ii}] & 10300,$^a$ 11200,$^d$ 13100,$^e$ 12400$^f$ & $11800\pm 1300$ & $12350\pm 1250^n$ \\

              &       & [S~{\sc ii}] & 12500,$^a$ 14100,$^g$ 15400$^h$          & $14000\pm 1500$ & $11050^{+3550n}_{-2050}$ \\

              &       & [O~{\sc iii}] & 8800,$^a$ 9100,$^b$ 9000,$^i$ 8700$^j$  & $8900\pm 200$  & $9060\pm 200^n$ \\

              &       & [S~{\sc iii}]  & 8800,$^a$ 9800,$^k$ 9500$^l$           & $9400\pm 500$  & $8800\pm 500^n$ \\

 Orion        & 8900  & [N~{\sc ii}]   & 10000,$^a$ 10900,$^b$ 10000$^c$        & $10300\pm 600$ & $10150\pm 350^v$ \\

              &       & [O~{\sc ii}]   & 7900,$^a$ 8400,$^d$ 9200,$^e$ 9000$^f$ & $8600\pm 600$  & $9800\pm 800^v$ \\

              &       & [S~{\sc ii}]   & 8700,$^a$ 9300,$^g$ 9800$^h$           & $9300\pm 600$  & $9050\pm 800^v$ \\

              &       & [O~{\sc iii}]  & 8000,$^a$ 8400,$^b$ 8300,$^i$ 7900$^j$ & $8200\pm 250$  & $8300\pm 40^v$ \\

              &       & [S~{\sc iii}]  & 9500,$^a$ 10700,$^k$ 10200$^l$         & $10100\pm 600$ & $10400^{+800t}_{-1200}$ \\

              &       & [Ar~{\sc iii}] & 8700,$^a$ 8300,$^k$ 9300$^m$           & $8700\pm 500$  & $8300\pm 400^v$ \\

HH\,202 (shock)& 17000& [N~{\sc ii}]   & 9700,$^a$ 9500,$^b$ 9800$^c$          & $9700\pm 150$  & $9240\pm 300^t$ \\

              &       & [O~{\sc ii}]   & 8500,$^a$ 9100,$^d$ 9900,$^e$ 9600$^f$ & $9300\pm 650$  & $9250\pm 280^t$ \\

              &       & [S~{\sc ii}]   & 9700,$^a$ 10400,$^g$ 11500$^h$         & $10500\pm 900$ & $8250\pm 540^t$ \\

              &       & [O~{\sc iii}]  & 8500,$^a$ 8800,$^b$ 8800,$^i$ 8400$^j$ & $8600\pm 200$  & $8770\pm 240^t$ \\
              
              &       & [S~{\sc iii}]  & 9100,$^a$ 10200,$^k$ 9900$^l$          & $9700\pm 600$  & $9280\pm 300^t$ \\

              &       & [Ar~{\sc iii}] & 8600,$^a$ 8200,$^k$ 9200$^m$           & $8700\pm 500$  & $8260\pm 410^t$ \\
\enddata
\tablenotetext{a}{Estimated with effective collision strengths from present work}
\tablenotetext{b}{Estimated with effective collision strengths from \citet{len94}}
\tablenotetext{c}{Estimated with effective collision strengths from \citet{tay11}}
\tablenotetext{d}{Estimated with effective collision strengths from \citet{pra06}}
\tablenotetext{e}{Estimated with effective collision strengths from \citet{tay07}}
\tablenotetext{f}{Estimated with effective collision strengths from \citet{kis09}}
\tablenotetext{g}{Estimated with effective collision strengths from \citet{cai93}}
\tablenotetext{h}{Estimated with effective collision strengths from \citet{tay10}}
\tablenotetext{i}{Estimated with effective collision strengths from \citet{agg99}}
\tablenotetext{j}{Estimated with effective collision strengths from \citet{pal12}}
\tablenotetext{k}{Estimated with effective collision strengths from \citet{gal95}}
\tablenotetext{l}{Estimated with effective collision strengths from \citet{hud12}}
\tablenotetext{m}{Estimated with effective collision strengths from \citet{mun09}}
\tablenotetext{n}{Estimated in the observational work of \citet{gar06}}
\tablenotetext{r}{Estimated in the observational work of \citet{gar05}}
\tablenotetext{s}{Estimated in the observational work of \citet{gar04}}
\tablenotetext{t}{Estimated in the observational work of \citet{mes09}}
\tablenotetext{v}{Estimated in the observational work of \citet{est04}}
\end{deluxetable}


\begin{deluxetable}{lrrrrr}
\tablewidth{0pt}
\tablecaption{$A$-value dependence of [\ion{S}{2}] temperature \label{Aval_temp}}
\tablecolumns{6}
\tablehead{\colhead{Source} & \colhead{$N_e$} & \colhead{$T_e^a$} & \colhead{$T_e^b$} & \colhead{$T_e^c$} & \colhead{$T_e^d$} \\
                            & \colhead{(cm$^{-3}$)} & \colhead{(K)} & \colhead{(K)} & \colhead{(K)}& \colhead{(K)}}
\startdata
M20           & 270   & 8800  & 9000  & 8900  & 8800  \\
S311          & 310   & 9300  & 9500  & 9400  & 9200  \\
M16           & 1100  & 9200  & 9500  & 9300  & 8700  \\
NGC~3576      & 2800  & 7600  & 7800  & 7700  & 7000  \\
HH~202 (neb)  & 2900  & 8300  & 8600  & 8400  & 7700  \\
NGC~3603      & 5200  & 12500 & 13000 & 12700 & 10700 \\
Orion         & 8900  & 8700  & 8900  & 8800  & 7700  \\
HH~202 (shock)& 17000 & 9700  & 10000 & 9800  & 8300  \\
\enddata
\tablenotetext{a}{Present work}
\tablenotetext{b}{Estimated with $A$-values from \citet{men82}}
\tablenotetext{c}{Estimated with $A$-values from \citet{fri99}}
\tablenotetext{d}{Estimated with $A$-values from \citet{iri05}}
\end{deluxetable}


\end{document}